\documentclass[showpacs,nofootinbib,amsmath,amssymb,superscriptaddress]{revtex4}
\usepackage[latin1]{inputenc}
\usepackage[T1]{fontenc}
\usepackage{graphicx}
\usepackage{epsfig}

\usepackage{dsfont}
\usepackage{amsmath,subfigure}
\usepackage{amssymb,braket}
\usepackage{hyperref}
\newcommand{\p}{\partial}

\begin{document}

\newcommand{\mhalfo}{\frac{1}{2}}	
\newcommand{\mhalf}[1]{\frac{#1}{2}}
\newcommand{\om}{\overline{m}}
\newcommand{\omu}{\overline{\mu}}
\newcommand{\ka}{\kappa}
\newcommand{\al}{\alpha}
\newcommand{\be}{\beta}
\newcommand{\ga}{\gamma}
\newcommand{\la}{\lambda}
\newcommand{\equ}[1]{\begin{equation} #1 \end{equation}}
\newcommand{\ali}[1]{\begin{align} #1 \end{align}}
\newcommand{\eref}[1]{eq.~(\ref{#1})}
\newcommand{\fref}[1]{fig.~\ref{#1}}
\newcommand{\ddotp}[1]{\frac{d^d #1}{(2\pi)^d}}	
\newcommand{\nnnl}{\nonumber\\}	
\newcommand{\G}[1]{\Gamma(#1)}
\newcommand{\nq}{\nu_1}	
\newcommand{\nw}{\nu_2}	
\newcommand{\nd}{\nu_3}	
\newcommand{\de}{\delta} 
\newcommand{\fig}[4]{\begin{figure}[#1]\centering\epsfig{file=#3}\caption{#4}\label{#2}\end{figure}}
\newcommand{\eps}[0]{\varepsilon} 
\newcommand{\pars}[0]{\partial^2} 
\newcommand{\vp}[0]{\varphi} 
\newcommand{\vpb}[0]{\bar{\varphi}} 
\newcommand{\vt}[0]{\vartheta}
\newcommand{\lap}[0]{-\partial^2}
\newcommand{\gam}[0]{\hat{\gamma}}
\newcommand{\spade}[0]{$\spadesuit$ }

\allowdisplaybreaks 

\title{Gribov horizon and $i$-particles: about a toy model and the construction of physical operators}

\author{L. Baulieu}\email{baulieu@lpthe.jussieu.fr}
\affiliation{Theory Division, CERN, 1211-Geneve 23, Switzerland}
\affiliation{LPTHE, Universit\'{e}s Pierre et Marie Curie, 4 place Jussieu, F-75252 Paris Cedex 05, France}

\author{D. Dudal}\email{david.dudal@ugent.be}
\affiliation{
Ghent University, Department of Physics and Astronomy, Krijgslaan 281-S9, B-9000 Gent, Belgium
}

\author{M. S. Guimaraes}
\email{marceloguima@gmail.com}
\affiliation{
UERJ - Universidade do Estado do Rio de Janeiro, Instituto de F\'isica - Departamento de F\'isica Te\'orica, Rua S\~ao Francisco Xavier 524, 20550-013 Maracan\~a,
Rio de Janeiro, Brasil
}

\author{M. Q. Huber}
\email{markus.huber@uni-jena.de}
\affiliation{Institut f\"ur Physik, Karl-Franzens-Universit\"at Graz, Universit\"atsplatz 5, 8010 Graz, Austria}

\author{S. P. Sorella}
\email{sorella@uerj.br}
\affiliation{
UERJ - Universidade do Estado do Rio de Janeiro, Instituto de F\'isica - Departamento de F\'isica Te\'orica, Rua S\~ao Francisco Xavier 524, 20550-013 Maracan\~a,
Rio de Janeiro, Brasil
}

\author{N. Vandersickel}\email{nele.vandersickel@ugent.be}
\affiliation{
Ghent University, Department of Physics and Astronomy, Krijgslaan 281-S9, B-9000 Gent, Belgium
}

\author{D. Zwanziger}\email{daniel.zwanziger@nyu.edu}
\affiliation{New York University, New York, NY 10003, USA}
\date{\today}

\begin{abstract}
\noindent Restricting the functional integral to the Gribov region $\Omega$ leads to a deep modification of the behavior of Euclidean Yang-Mills theories in the infrared region. For example, a gluon propagator of the Gribov type,
$\frac{k^2}{k^4+{\hat \gamma}^4}$, can be viewed as  a propagating pair of unphysical modes, called here $i$-particles, with complex masses $\pm i{\hat \gamma}^2$. From this viewpoint, gluons are unphysical and one can see them as being confined.  We introduce a simple toy model describing how a suitable set of composite operators can be constructed out of $i$-particles whose correlation functions exhibit only real branch cuts, with associated positive spectral density. These composite operators can thus be called physical  and are the toy analogy of glueballs in the Gribov-Zwanziger theory.

\end{abstract}

\pacs{11.10.-z, 03.70.+k, 11.15.-q}

\maketitle

\section{Introduction}\label{sec:intro}
A very peculiar feature of the strong interaction is the fact that
quarks and gluons cannot be observed as free particles, {\it
i.e.}~they are absent from the physical spectrum. As the theory
of quantum chromodynamics (QCD) describes the strong interaction,
it should account for this property. A first step towards a full
picture is the investigation of the gluonic sector alone, thus
ignoring contributions from quarks. Even this simpler case of
Yang-Mills theory is very intricate and is widely studied
from both analytical and numerical points of view. Its spectrum is believed to consist of so-called
glueballs, color singlets made up by gluons only\footnote{For a recent review on glueballs, see \cite{Mathieu:2008me}.}. Yang-Mills theory then has to explain why gluons do not exist as asymptotic
states, whereas glueballs do. \\\\A possible and well investigated
mechanism for gluon confinement is provided by the
Gribov-Zwanziger framework  \cite{Gribov:1977wm,Zwanziger:1989mf}, which enables us to take into account the
effect of Gribov copies  \cite{Gribov:1977wm}. Although the existence of Gribov copies is a general feature
of the gauge fixing procedure \cite{Singer:1978dk}, we shall refer here to the case of the
Landau gauge, $\partial_{\mu}A^a_{\mu}=0$,
which is a renormalizable gauge in the continuum, while also possessing a lattice
formulation. The problem of gauge copies is remedied
by restricting the domain of integration
in the Feynman path integral to the Gribov region $\Omega$, defined as the set of field
configurations which obey the Landau condition and for which the Faddeev-Popov operator, ${\cal M}^{ab}=-\partial_\mu D_\mu^{ab}  = -(\delta^{ab}\partial^2+g\,f^{abc} A_\mu^c \partial_\mu)$,  is strictly positive, namely
\begin{align}
\Omega = \{ \; A; \; \partial_{\mu} A^a_{\mu}=0, \; {\cal M}^{ab} > 0 \; \}   \;.  \label{om}
\end{align}
The region $\Omega$ is convex and bounded in all directions in field space, and every gauge orbit passes through $\Omega$ \cite{Dell'Antonio:1991xt}. Its boundary $\partial \Omega$ is known as the first Gribov horizon, where the  lowest eigenvalue of the Faddeev-Popov operator vanishes. As shown in \cite{Zwanziger:1989mf, Zwanziger:1993dh}, the restriction to the region $\Omega$ is achieved by adding to the Yang-Mills action the so-called horizon term
\begin{align}
S_h = \gamma^4 \int d^4 x \,h(x) = \gamma^4  \int d^4 x \int  d^4 y  D_\mu^{ac}(x) (\mathcal M^{-1})^{ad}(x,y) D_\mu^{dc} (y) \;,   \label{ho}
\end{align}
with
\begin{eqnarray}
D_\mu^{ab} &=& \p_\mu \delta^{ab} +  g f^{acb} A^c_\mu \;,
\end{eqnarray}
the covariant derivative.  The action thus becomes
\begin{align}
S =S_{YM}+S_{gf}+S_h \;, \label{actGZ1}
\end{align}
where the Yang-Mills and gauge fixing part of $S$ are given by
\begin{align}
S_{YM}&=\frac1{4}\int d^4 x F_{\mu\nu}^{a}F_{\mu\nu}^{a} \;,\\
S_{gf}&=\int d^4 x\;  (i\,b^a \partial_\mu A_\mu^a-\bar{c}^a {\cal M}^{ab} c^b) \; . \label{act2}
\end{align}
The field $b^a$ is the Lagrange multiplier enforcing the Landau gauge condition, and $({\bar c}^a, c^a)$ are the Faddeev-Popov ghosts. The massive parameter $\gamma$, known as the Gribov parameter,  is not free, being determined in a self-consistent way by a gap equation called the
horizon condition \cite{Zwanziger:1989mf}, namely
\begin{align}
\label{gap-non-local}
\langle h(x) \rangle =d (N^2-1) \; ,
\end{align}
where $N$ is the number of colors and $d$ the space-time dimension. \\\\Although being nonlocal, the horizon term can be cast in local form by introducing a suitable set of auxiliary fields $(\bar{\vp}_\mu^{ab}, \vp_\mu^{ab}, \omega_\mu^{ab}, \bar{\omega}_\mu^{ab})$. The fields $(\bar{\vp}_\mu^{ab}, \vp_\mu^{ab})$ are a pair of complex conjugate bosonic fields,  whereas  $(\omega_\mu^{ab}, \bar{\omega}_\mu^{ab})$ are anticommuting.
Notice that these fields carry two color indices and one Lorentz index. This is necessary in order to account for the required number  of degrees of freedom. Thus, the localized form of the action reads
\begin{align}
S_{GZ}&=S_{YM}+S_{gf}+S_{loc} \;, \label{locact1}
\end{align}
where
\begin{align}
S_{loc}&=\int d^4x\left( \bar{\vp}_\mu^{ac} {\cal M}^{ab} \vp_\mu^{bc}-\bar{\omega}_\mu^{ac} {\cal M}^{ab} \omega_\mu^{bc}
-g\,f^{abc}(\partial_\nu\bar{\omega}_\mu^{ad})(D^{be}_\nu c^e)\vp_\mu^{cd}
+\gamma^2\,g\,f^{abc}A_\mu^{abc}(\vp_\mu^{bc}-\bar{\vp}_\mu^{bc})-d(N^2-1)\gamma^4 \right) \;. \label{locact12}
\end{align}
The last term is introduced in order to be able to rewrite the horizon condition (\ref{gap-non-local}) as
\begin{align}
\frac{\delta\Gamma}{\delta \gamma^2}=0 \;,  \label{geq}
\end{align}
where $\Gamma$ is the vacuum energy, {\it i.e.}
\begin{equation}
e^{-\Gamma} = \int [d\Phi] \; e^{-S_{GZ}} \;,  \label{vc}
\end{equation}
and $[d\Phi]$ stands for the functional integration over all fields. In terms of the auxiliary fields $(\bar{\vp}_\mu^{ab}, \vp_\mu^{ab})$, the horizon condition (\ref{geq}) takes the form
\begin{equation}
  \left\langle  gf^{abc} A^a_{\mu} (\vp_\mu^{bc}-\bar{\vp}_\mu^{bc})  \right\rangle = 2 d (N^2-1) \gamma^2 \;. \label{gep1}
\end{equation}
Remarkably, the action ({\ref{locact1}) turns out to be multiplicatively renormalizable to all orders, thanks to the existence of a rich set of Ward identities which can be established at the quantum level \cite{Zwanziger:1989mf,Zwanziger:1992qr,Maggiore:1993wq,Dudal:2005na,Gracey:2005cx,Gracey:2006dr,Dudal:2008sp}. Let us mention  in particular that, unlike the case of the Faddeev-Popov action, the Gribov-Zwanziger action does not enjoy exact BRST symmetry, which turns out to be softly broken by terms proportional to the Gribov parameter $\gamma$ \cite{Dudal:2008sp,Baulieu:2008fy,Sorella:2009vt}. Nevertheless, the breaking term can be kept under control at the quantum level, giving rise  to softly broken Slavnov-Taylor identities which constrain very much the form of the possible counterterm. We also refer to \cite{Dudal:2010fq} for a few peculiarities concerning the renormalization analysis of the Gribov-Zwanziger theory. \\\\The restriction to the Gribov region $\Omega$ entails deep modifications  in the behavior of the theory in the infrared region. Let us give a look, for example, at the gluon propagator stemming from the action \eqref{locact1}
\begin{equation}
\left\langle A_{\mu }^{a}(k)A_{\nu }^{b}(-k)\right\rangle =\delta
^{ab}\left( \delta _{\mu \nu }-\frac{k_{\mu }k_{\nu }}{k^{2}}\right) \frac{%
k^{2}}{k^{4}+{\hat \gamma }^{4}}\;, \qquad {\hat{\gamma}}^4 = 2 g^2 N \gamma^4 \;. \label{z11}
\end{equation}%
Expression \eqref{z11}  is suppressed in the infrared region,
while displaying complex poles at
$k^{2}=\pm i{\hat \gamma}^{2}$. This structure does not  allow
us  to attach the
usual particle meaning to the gluon propagator, invalidating the
interpretation of gluons as excitations of the physical spectrum.
In other words,  gluons cannot be considered as
part of the physical spectrum. In this sense, they are confined by the Gribov horizon, whose presence is encoded in the explicit dependence of expression \eqref{z11}  on
the Gribov parameter $\gamma$. Another way to see that  gluons are not  physical particles is the observation that the propagator (\ref{z11}) has negative norm contributions and thus it violates positivity \cite{Alkofer:2003jj}. Since a positive-definite norm is required for a probabilistic interpretation \cite{Osterwalder:1973dx,Osterwalder:1974tc}, we conclude that the excitations described by \eref{z11} are confined. One might argue that further nonperturbative effects besides the Gribov horizon might conspire to change this picture. However, qualitatively one can obtain the same result with functional methods \cite{vonSmekal:1997is,vonSmekal:1997vx,Zwanziger:2001kw,Lerche:2002ep,Pawlowski:2003hq,Huber:2009tx}. \\\\Even if the restriction to the Gribov region, with the consequence of an infrared  soft breaking of the BRST symmetry, provides a natural mechanism for gluon confinement,  one is left with the   nontrivial task of constructing a set of physical composite operators made of
the fields entering the Gribov-Zwanziger action  (\ref{locact1}),  whose correlation functions can be given a physical meaning in Minkowskian space.  A possible strategy to work out this kind of analysis is to look at the spectral properties of the correlation functions. In practice, this amounts to face the challenging task of constructing a suitable set of composite operators whose correlation functions can be cast in the form of a K\"all\'{e}n-Lehmann  representation, {\it i.e.} a spectral representation  with a positive spectral function, and whose analytic continuation in the complex Euclidean $k^2$-plane exhibits  a cut along the negative real axis only. Such a spectral representation would thus imply that, when moving to Minkowskian space, the cuts are
located along the positive real axis. Moreover, positivity of the spectral function then guarantees
that a meaningful interpretation in terms of states of a physical spectrum can be attached to those  operators.  This is precisely what one would expect from a confining theory. This is a highly nontrivial task, given the complexity of the Gribov type propagator \eqref{z11} as well as of the Gribov-Zwanziger action \eqref{locact1}. \\\\The first step towards a detailed study of the analyticity properties of the correlation functions within the Gribov-Zwanziger framework was undertaken by one of the authors \cite{Zwanziger:1989mf}, who evaluated at one loop order the correlation function of two gauge invariant operators, namely
\begin{equation}
G(k^2)=\int d^{4}x\ e^{-ikx\ }\left\langle F^{2}(x)F^{2}(0)\right\rangle \; ,
\label{z2}
\end{equation}
where $F^2(x)=F^a_{\mu\nu}(x) F^a_{\mu\nu}(x)$.  The results found in \cite{Zwanziger:1989mf} can be summarized as follows:
\begin{equation}
G(k^2) = G^{\rm phys}(k^2) + G^{\rm unphys}(k^2) \;, \label{z3}
\end{equation}
The unphysical part, $G^{\rm unphys}(k^2)$, displays cuts along the imaginary  axes beginning at the unphysical values $k^2=\pm 4i{\hat \gamma}^2$, whereas the
physical part, $G^{\rm phys}(k^2)$, has a cut beginning at the
physical threshold $k^2=-2{\hat \gamma}^2$. Moreover, the spectral
function of  $G^{\rm phys}(k^2)$ turns out to be positive, so that it  possesses a
K\"all\'{e}n-Lehmann representation \cite{Zwanziger:1989mf}.
As such, $G^{\rm phys}(k^2)$ is an acceptable correlation function for
physical glueball excitations. What is interesting in
expression \eqref{z3} is that a physical cut has emerged in the
correlation function of  a gauge invariant quantity,
even if it has
been evaluated with a gluon propagator exhibiting only unphysical
complex poles.\\\\ Obviously, the big challenge lying ahead of us is to find out if one can get rid of the unwanted unphysical cut. We should like to mention an alternative possibility which at the moment remains at the level of a scenario, in fact, a scenario with a deus ex machina:  it is possible that physically satisfactory analyticity properties emerge only when the horizon condition, eq.(\ref{geq}), is satisfied.  This is difficult to investigate because the horizon condition is nonperturbative, and we can offer no evidence in support of this happy eventuality.  In the present investigation we are limited to perturbative calculations.   \\\\The present paper aims at pursuing such an analysis. To go further, we should figure out a plausible mechanism which deforms local gauge invariant operators $O_i$ such that their expectation values have only physical cuts. This deformation can be done by adding to $O_i$ local quantities depending on the fields of the Gribov-Zwanziger action. As it  will be  shown, this deforming mechanism will allow, in particular, to get rid of the unphysical cuts in the lowest non trivial order for the  above-mentioned two point function, eq.\eqref{z2}.  The possibility that the mechanism can be generalized  for other operators, and at any given order of perturbation theory, could be envisaged as follows. One first considers the usual observables of Yang-Mills theory: basically the elements of the BRST cohomology, {\it i.e.} the color singlet local gauge invariant operators built up of the field strength $F^a_{\mu\nu}$ and its covariant derivatives. For a given canonical dimension, one has generally a mixing that is induced by renormalization \cite{collins,Dudal:2009zh}. For instance,  an operator  such as  $F^3= f^{abc} F^{a}_{\mu\nu}F^b_{\nu\rho} F^c_{\rho\mu}$ can mix with other BRST nontrivial operators such as $F^a_{\mu\nu}(D^2F_{\mu\nu})^a$,  etc.  Moreover, the softly broken Slavnov-Taylor identities indicate that, besides the expected mixing with BRST exact operators, also mixings with operators which are not BRST invariant will, in general, occur.  One  hopes,  however, that only suitable  combinations of all these operators can be defined, such that their expectation values only exhibit physical cuts,  with mutual compensation of all terms with unphysical cuts.  One is thus led to a kind of boot-strap mechanism, where one starts with a particular mixing at the tree level, such that the absence of unphysical cuts be enforced order by order in perturbation theory. In other words, at the tree  level, an  observable should be written  in the following form
\begin{align}
{\cal O} =\sum_i  \left( \;  \alpha^i{(g^2) } {\cal O}^{\rm ( BRST \ {invariant})}_i + \beta^i(g^2)  {\cal O}^{\rm (non\ BRST\ {invariant})}_i  \; \right)
\;. \label{laurent}
\end{align}
The values of the coefficients $ \alpha^i(0)$  and  $ \beta^i(0)$ at their lowest order are  supposed to be determined by the requirement of having only physical cuts. Our point of view is thus as follows.  The BRST symmetry characterizes the action and  the observables of the theory in the perturbative short distance regime as elements of its cohomology.  This symmetry is however  softly broken  in the long distance regime. Even if the presence of the soft breaking  enables us to determine a renormalizable local action under the form of the Gribov-Zwanziger action, it turns out that the local  observables have to  be modified  by the addition of suitable terms as in eq.(\ref{laurent}), where the positivity of the corresponding correlation functions is required to hold. When all the auxiliary fields are integrated out, this is tantamount to  modifying the observables and the action by nonlocal terms, proportional to the nonperturbative  Gribov parameter $\gamma$.  The remaining of this paper will consist in providing a certain number of examples in order to check the plausibility of this framework, by using the predictive power of local quantum field theory. \\\\Our main purpose here is therefore  that of investigating how the use of a confining Gribov type propagator can allow us to obtain examples of correlation functions exhibiting real cuts only and positive spectral functions. Our analysis relies on the introduction of  what we shall call $i$-particles: a pair of fields with
complex conjugate masses which emerge in a natural way when dealing with a Gribov type propagator. In fact
\begin{align}
\frac{1}{k^4+{\hat \gamma}^4} = \frac{1}{2i {\hat \gamma}^2} \left( \frac{1}{k^2-i{\hat \gamma}^2} - \frac{1}{k^2+i {\hat \gamma}^2 }\right)  \;,
\end{align}
from which it is apparent that a Gribov  propagator can be associated to the propagation of excitations with complex conjugate masses $\pm i{\hat \gamma}^2$. As it will be discussed in details, the use of $i$-particles enables us to provide examples of operators whose correlation functions exhibit the desired analyticity properties. More specifically, we shall be able to show that local composite operators built up with pairs of $i$-particles display cuts along the negative real axis in the complex Euclidean $k^2$-plane, while giving rise to positive spectral functions. \\\\
The paper is organized as follows. In Sections \ref{sec:model1}-\ref{i-fields} we discuss our ideas with a toy model, which is introduced in Section \ref{sec:model1} using a scalar field possessing a confining  propagator of the Gribov type. This will enable us to illustrate in a simple way how the action of the toy model can be cast in diagonal form through the introduction of $i$-particles.  Section \ref{sec:cmpop1} is devoted to a detailed study of a first example of a local operator, constructed from $i$-particles, whose one loop correlation function  possesses a  K\"all\'{e}n-Lehmann spectral representation. Section \ref{details} provides further details of the calculation of this correlation function in the complex $k^2$-plane.  In Section \ref{higherloop} we face the evaluation of correlation functions  with several loops.  The evaluation of the spectral representation for  two and three loop integrals obtained from operators built up with $i$-particles will be worked out. In particular, it will be shown that the spectral function at higher order can be obtained by a kind of convolution of the spectral functions obtained at  lower orders. This provides a useful iterative procedure, allowing us to generalize the argument to the $n$-loop case.  In Section \ref{i-fields} we discuss the meaning of the $i$-fields within the toy model. We argue that those fields might be regarded as rather useful variables in order to construct sensible composite operators within a local quantum field theory framework. In Section \ref{gz} we return to the Gribov-Zwanziger action. We discuss how $i$-particles emerge and how they can be employed to give examples of one loop correlation functions which, unlike expression (\ref{z3}), display cuts along the negative real axis only while having positive spectral functions. Even if a systematic construction of a set of meaningful operators in the presence of the Gribov horizon is far form being realized, the introduction of the $i$-particles might be seen as an interesting path in order to provide at least some examples of operators displaying good analyticity properties, something which can already be regarded as a nontrivial achievement. In the Conclusions a few points for future  investigations will be spotted. Finally, several useful details about the evaluation of the spectral densities are reported in the Appendices.

\section{A scalar field theory toy model}\label{sec:model1}
\subsection{ Constructing the toy model}
A simple way of constructing a field theory model exhibiting a confining Gribov type propagator is through a scalar field $\psi$ whose  Euclidean action is nonlocal, being specified by
\begin{align}
S = \int d^4x \; \frac{1}{2}  \psi \left( -\partial^2 + 2\frac{\theta^4}{-\partial^2} \right) \psi \;. \label{act}
\end{align}
The massive parameter $\theta$ is introduced by hand and is akin to  the Gribov parameter $\gamma$.  As it is easily seen, the resulting propagator is in fact of the Gribov type
\begin{align}
\langle \psi(k) \psi(-k) \rangle = \frac{k^2}{k^4+2\theta^4} \;. \label{gp}
\end{align}
Although being nonlocal, expression (\ref{act}) can be cast in local form by introducing a pair of bosonic complex conjugate fields $({\bar \varphi}, \varphi)$ and a pair of anticommuting fields $({\bar \omega}, \omega)$, so that
\begin{align}
S =  \int d^4x \; \left( \; \frac{1}{2}  \psi ( -\partial^2 ) \psi +  {\bar \varphi} (-\partial^2) \varphi +  {\theta^2}\psi (\varphi-\bar{\varphi})
- {\bar \omega}(-\partial^2) \omega   \right) \;.  \label{lact}
\end{align}
Let us have a look at the propagators of the model. To that purpose, we introduce new variables $(U,V)$ in order to achieve a partial diagonalization of the action
\begin{align}
\varphi  & = \frac{U+iV}{\sqrt{2}} \;,  \nonumber \\
{\bar \varphi}  & = \frac{U-iV}{\sqrt{2}} \; . \label{v1}
\end{align}
In terms of the fields $(U,V)$, the action of the toy model reads
\begin{equation}
S = \int d^4 x \; \left( \; \frac{1}{2} \psi (-\partial^2) \psi + \frac{1}{2} V (-\partial^2) V + \sqrt{2} i \theta^2 \psi V
+ \frac{1}{2} U (-\partial^2) U - {\bar \omega} (-\partial^2) \omega \; \right) \;. \label{act1}
\end{equation}
From the above expression, the propagators can be worked out:
\begin{align}
\langle \psi \psi \rangle & = \frac{p^2}{p^4+2\theta^4} \;, \nonumber \\
\langle V V \rangle & =     \frac{p^2}{p^4+2\theta^4} \;, \nonumber \\
\langle V \psi \rangle & =  \frac{-i\sqrt{2} \theta^2}{p^4+2\theta^4} \;, \nonumber \\
\langle U U \rangle & = \frac{1}{p^2} \;. \label{prop}
\end{align}
Having evaluated the propagators of the fields $(\psi, U, V)$, we can now check what  the propagators in terms
of the fields $(\varphi, {\bar \varphi})$ are.
One finds
\begin{align}
\langle \psi \varphi \rangle & = \frac{\theta^2}{p^4 + 2 \theta^4} \;, \nonumber \\
\langle \psi {\bar \varphi} \rangle & = - \frac{\theta^2}{p^4 + 2 \theta^4} \;, \nonumber \\
\langle  \varphi {\bar \varphi} \rangle & = \frac{p^4 +\theta^4}{p^2(p^4 + 2 \theta^4)} \;, \nonumber \\
\langle \varphi \varphi \rangle & = \frac{\theta^4}{p^2(p^4 + 2 \theta^4)} \;.  \label{np}
\end{align}

\subsection{Introducing the $i$-particles}\label{ip}
Expression (\ref{act1}) can be cast in complete diagonal form by making a further change of variables:
\begin{align}
\psi & = \frac{1}{\sqrt{2}} (\lambda +\eta) \;, \\
V & = \frac{1}{\sqrt{2}} (\lambda -\eta) \;, \label{le}
\end{align}
so that
\begin{equation}
S = \int d^4 x \; \left( \; \frac{1}{2} \lambda (-\partial^2+i\sqrt{2}\theta^2) \lambda + \frac{1}{2} \eta (-\partial^2-i\sqrt{2}\theta^2) \eta
+ \frac{1}{2} U (-\partial^2) U - {\bar \omega} (-\partial^2) \omega \; \right) \;. \label{actGZ2}
\end{equation}
From this expression one immediately sees that the fields $\lambda$ and $\eta$ correspond to the propagation of unphysical modes with complex masses $\pm i \sqrt{2} \theta^2$. These are the $i$-particles of the model, namely\footnote{There are more
propagators than the ones shown, but these are not relevant for the
calculation presented.}
\begin{align}
\langle \lambda(k) \lambda(-k) \rangle & = \frac{1}{k^2+i\sqrt{2}\theta^2} \; \\
 \langle \eta({k}) \eta({-k}) \rangle & = \frac{1}{k^2-i\sqrt{2}\theta^2} \;. \label{iprop}
 \end{align}
It is important to notice here that requiring
\begin{equation}\label{herm}
    \lambda^\dagger=-\eta\,,\qquad \eta^\dagger=-\lambda
\end{equation}
will ensure that \eqref{actGZ2} corresponds to a Hermitian action.

\section{ Construction of composite operators at one loop}\label{sec:cmpop1}
\subsection{Preliminaries}
Having cast the action in diagonal form, we face now the construction of a set of composite operators made out of
$i$-particles, whose correlation functions exhibit real cuts only. As the Gribov propagator (\ref{gp}) and the diagonal form of the action (\ref{act2}) suggest, we shall look at composite operators constructed by means of pairs of $i$-particles.\\\\
The simplest example which one can consider  at one loop is that of the dimension two composite operator  consisting of
one $i$-particle of the type $\lambda$ and one $i$-particle of the type $\eta$, namely
\begin{align}
O_1(x) = \lambda(x) \eta(x) \;, \label{o1}
\end{align}
which is indeed a real (Hermitian) operator, according to the prescription \eqref{herm}.\\\\The correlation function $\langle O_1(k) O_1(-k) \rangle $
in $d$ Euclidean dimensions is given by \begin{align}
\langle O_1(k) O_1(-k) \rangle = \int \frac{d^dp}{(2\pi)^d}\; \frac{1}{(k-p)^2-i\sqrt{2}\theta^2} \frac{1}{p^2+i\sqrt{2}\theta^2} \;. \label{o1c}
\end{align}
By direct inspection of the action (\ref{act2}), it follows that the  correlation function of three operators $O_1(x)$ vanishes
\begin{align}
\langle O_1(x) O_1(y) O_1(z) \rangle = 0 \;. \label{v}
\end{align}
Only correlation functions with an even number of operators $O_1$ are nonvanishing.\\\\The complex nature of the $i$-particle masses may raise concerns about the consistency of the theory. A pressing question is that of the unitarity. One has to be sure that these correlation functions have a well defined probabilistic interpretation. In order to settle this question, in what follows we shall provide an explicit  evaluation of the spectral function associated with the K\"all\'{e}n-Lehmann representation of the two-point function $\langle O_1(k) O_1(-k) \rangle$.

\subsection{K\"all\'{e}n-Lehmann representation in $d=4$}\label{subsubsec:D4}
We start our analysis by evaluating the integral (\ref{o1c}). Our way to proceed is that of casting it into the so-called spectral representation, which has a precise and powerful meaning, both in complex analysis and quantum field theory.  It is given by
\begin{align}
 \langle O_1(k) O_1(-k) \rangle  =  \int_{\tau_{0}}^{\infty} d\tau \; \rho({\tau}) \; \frac{1}{\tau+k^2} \;, \label{a3-1}
\end{align}
where the quantity $\tau_{0}>0$ stands for the threshold. Once expression (\ref{a3-1}) has been obtained, we take it as our very starting point. We introduce the complex function
\begin{align}
F(z) =    \int_{\tau_{0}}^{\infty} d\tau \; \rho(\tau) \; \frac{1}{\tau+z} \;. \label{a4}
\end{align}
Therefore, from complex analysis, it follows that $F(z)$ is an analytic function in the cut complex plane, where the interval $(-\infty, -\tau_{0})$ has been excluded. Moreover, when moving from Euclidean to Minkowskian space, {\it i.e. $k^2_{Eucl} \rightarrow - k^2_{Mink}$}, expression (\ref{a4}) gives the spectral representation of a quantity exhibiting a discontinuity along the positive real axis, starting at the threshold $\tau_{0}$ and extending till $+\infty$.  One requires also that the function $\rho(\tau)$ is positive in order to have a probabilistic interpretation. Indeed, using Cauchy's theorem, it is a relatively easy exercise to show that
\begin{equation}\label{16}
\rho(\tau)=\frac{1}{2\pi i}\lim_{\epsilon\to 0^+}\left[F(-\tau-i\epsilon)-F(-\tau+i\epsilon)\right]\,,\qquad\textrm{with }\tau\geq\tau_0\,,
\end{equation}
{\it i.e.}~ $\rho(\tau)$ is directly related to the discontinuity of the propagator $F(z)$ along its branch cut. Via the optical theorem, we then also know that $\rho(\tau)$ is proportional to a cross section, which evidently needs to be positive in a meaningful quantum field theory.
\\\\In order to obtain the spectral representation, we  employ Feynman parameters in the general $d$-dimensional situation in order to combine the denominators in (\ref{o1c})
\begin{align}
\langle O_1(k) O_1(-k) \rangle &= \int \frac{d^dp}{(2\pi)^d}\; \frac{1}{(k-p)^2-i\sqrt{2}\theta^2} \frac{1}{p^2+i\sqrt{2}\theta^2}\nonumber\\
&= \int \frac{d^dp}{(2\pi)^d}\; \int^{1}_{0} dx \frac{1}{\left[x(k^2-2p\cdot k-2i\sqrt{2}\theta^2) + p^2 + i\sqrt{2}\theta^2\right]^2}\nonumber\\
&= \int \frac{d^dq}{(2\pi)^d}\; \int^{1}_{0} dx \frac{1}{\left[q^2 + (x-x^2)k^2 - (2x-1)i\sqrt{2}\theta^2 \right]^2}
\;, \label{fey-par}
\end{align}
where we have defined $q\equiv p-kx$. Using now the identity
\begin{align}
\int \frac{d^dq}{(2\pi)^d}\;\frac{1}{\left[q^2 + \Delta^2\right]^n} = \frac{1}{(4\pi)^{\frac{d}{2}}} \frac{\Gamma(n-\frac{d}{2})}{\Gamma(n)}(\Delta^2)^{\frac{d}{2}-n},
\label{fey-idty}
\end{align}
with $n=2$ and $\Delta^2 = (x-x^2)k^2 - (2x-1)i\sqrt{2}\theta^2$, we obtain
\begin{align}
\langle O_1(k) O_1(-k) \rangle &=  \frac{\Gamma(2-\frac{d}{2})}{(4\pi)^{\frac{d}{2}}} \int^{1}_{0} dx \left[(x-x^2)k^2 - (2x-1)i\sqrt{2}\theta^2 \right]^{\frac{d}{2}-2}
\;. \label{fey-par2}
\end{align}
We now consider the case of $d=4$.  We define
\begin{equation}
F(k^2)=\langle O_1(k) O_1(-k) \rangle\,.
\end{equation}
We start from \eqref{fey-par2} and act on it with $\frac{\p}{\p k^2}$. This  regularizes the original integral for $F(k^2)$, which is ultraviolet divergent.
After setting $d=4$, we find
\begin{equation}\label{int9}
\frac{\p}{\p k^2}F(k^2)= -\frac{1}{16\pi^2}\int_0^1 dx\frac{x(1-x)}{x(1-x)k^2+xi-i/2}=-\frac{1}{16\pi^2}\int_0^1 dx\frac{x(1-x)}{x(1-x)k^2-xi+i/2}\,,
\end{equation}
where we temporarily switched to units $2\sqrt{2}\theta^2=1$ for notational convenience. The substitution $s=\frac{2x-1}{2x(1-x)}$, or $x=\frac{-1+s+\sqrt{1+s^2}}{2s}$, brings us to
\begin{eqnarray}\label{int10}
\frac{\p}{\p k^2}F(k^2)&=& -\frac{1}{16\pi^2}\int_{-\infty}^{+\infty} ds \frac{1}{k^2-is} \frac{d\left(\frac{-1+s+\sqrt{1+s^2}}{2s}\right)}{ds}
\nonumber\\&=&\frac{1}{16\pi^2}\int_{-\infty}^{+\infty} \frac{ids}{(k^2-is)^2}\frac{-1+s+\sqrt{1+s^2}}{2s}\,,
\end{eqnarray}
where we employed partial integration. There is no problem at $s=0$ if $k^2>0$, since $\lim_{s\to0}\frac{-1+s+\sqrt{1+s^2}}{2s}=\frac{1}{2}$. There are no poles in the upper half $s$-plane for $k^2>0$, so we can deform the contour to be located around the cut for $s\in[i\infty,i]$. Setting $s=i\tau$, we compute \eqref{int10} as
\begin{eqnarray}\label{int10b}
\frac{\p}{\p k^2}F(k^2)&=& \frac{1}{16\pi^2}\int_{\infty}^{1}\frac{-d\tau}{(k^2+\tau)^2}\frac{-1+i\tau-i\sqrt{\tau^2-1}}{2i\tau}+\frac{1}{16\pi^2}\int^{\infty}_{1}\frac{-d\tau}{(k^2+\tau)^2}\frac{-1+i\tau+i\sqrt{\tau^2-1}}{2i\tau}\nonumber\\
&=&-\frac{1}{16\pi^2}\int^{\infty}_{1}\frac{\sqrt{\tau^2-1}}{\tau}\frac{d\tau}{(k^2+\tau)^2}\,.
\end{eqnarray}
We can subsequently integrate this expression from $0$ to $k^2$, finding
\begin{eqnarray}\label{int11}
F(k^2)-F(0)=\frac{1}{16\pi^2}\int_1^{\infty}\frac{\sqrt{\tau^2-1}}{\tau}\left(\frac{1}{\tau+k^2}-\frac{1}{\tau}\right)d\tau\,,
\end{eqnarray}
or, by restoring the units,
\begin{eqnarray}\label{int11b}
F(k^2)-F(0)=\frac{1}{16\pi^2}\int_{2\sqrt{2}\theta^2}^{\infty}\frac{\sqrt{\tau^2-8\theta^4}}{\tau}\left(\frac{1}{\tau+k^2}-\frac{1}{\tau}\right)d\tau\,.
\end{eqnarray}
From this expression, we notice the importance of the subtraction of $F(0)$ to find a finite result, otherwise we would find a divergent spectral integral. The spectral density can be read off from \eqref{int11b},
\begin{equation}\label{int12b}
\rho(\tau)=\frac{1}{16\pi^2}\frac{\sqrt{\tau^2-8\theta^4}}{\tau}\,.
\end{equation}
Interpreting this result as an expression of the physical spectrum of the theory, we see that it has a threshold at $2\sqrt{2}\theta^2$. The same calculation can be repeated in the case of two particles with real mass $\mu$,  where the spectrum is found to begin at the threshold $4\mu^2 = (\mu + \mu)^2$. In the present case we have an analogous situation for the $i$-particles, even though they have complex masses: $2\sqrt{2}\theta^2 = \left(e^{i\frac{\pi}{4}} 2^{\frac 14} \theta + e^{-i\frac{\pi}{4}} 2^{\frac 14} \theta \right)^2$. Note also that the spectral function is positive in the range of integration. This is a very important feature for the theory to have a physically meaningful probabilistic interpretation.\\\\Finally, an explicit integration of \eqref{int11} leads to
\begin{eqnarray}\label{int12}
F(k^2)-F(0)=\frac{1}{16\pi^2}\left(1-\frac{\pi}{2k^2}+\frac{\sqrt{1-k^4}}{k^2}\mathrm{arccos}(k^2)\right)\,.
\end{eqnarray}
In FIG.~\ref{fig1a} and \ref{fig1b}, we have displayed the (rescaled) real and imaginary part of $F(k^2)$. The cut for $z\in[-\infty,-1]$ is clearly visible.
\begin{figure}[h]
  \begin{center}
  \subfigure[]{\includegraphics[width=8.cm]{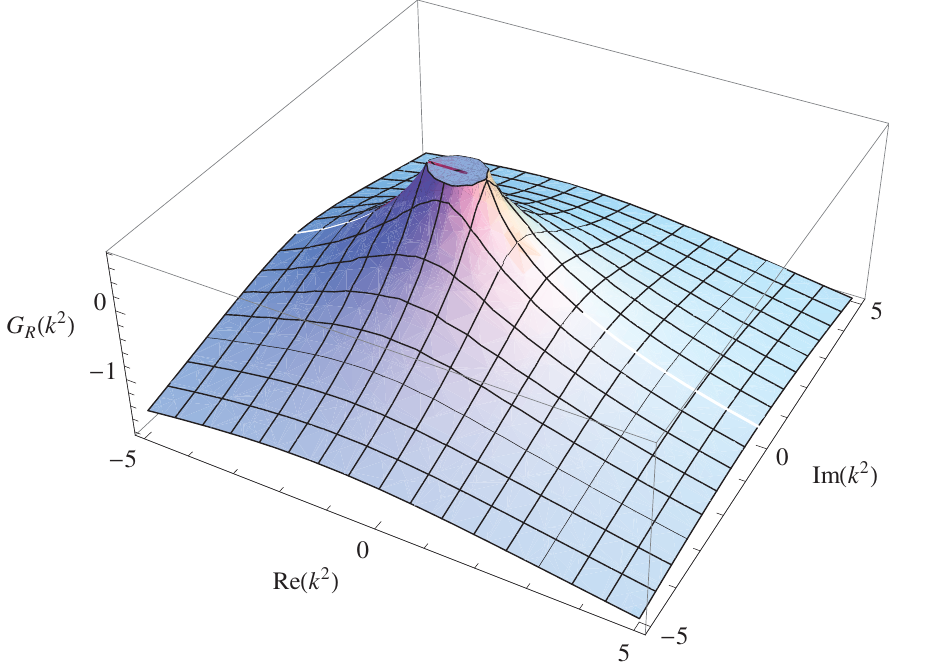} \label{fig1a}}
    \hspace{1cm}
    \subfigure[]{\includegraphics[width=8.cm]{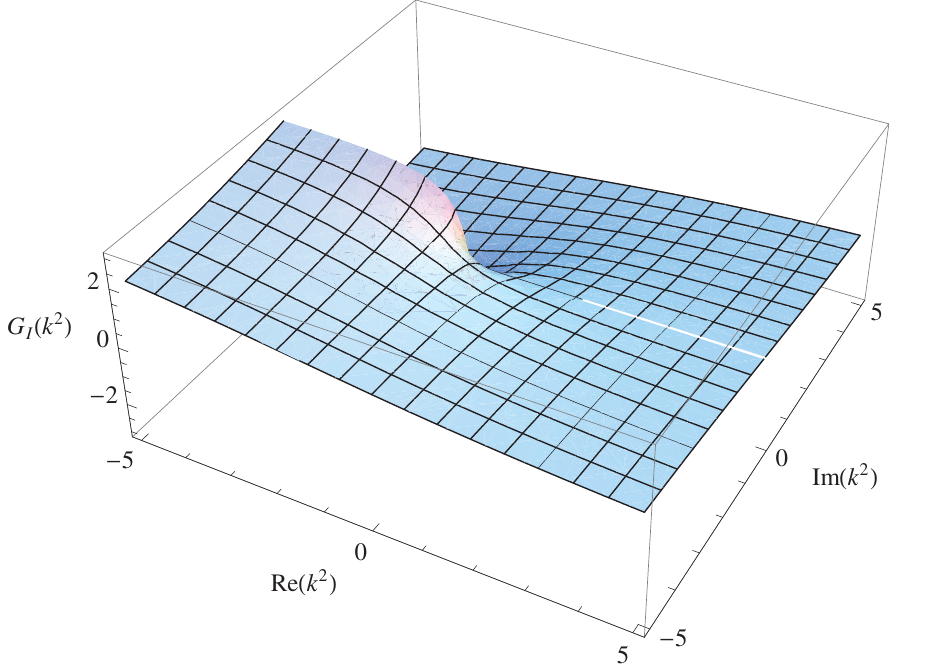}\label{fig1b}}
     \end{center}
  \caption{Plots of $G_R(k^2)\equiv16\pi^2 \mathrm{Re}[F(k^2)-F(0)]$ and $G_I(k^2)\equiv16\pi^2 \mathrm{Im}[F(k^2)-F(0)]$, respectively, with $F(k^2)-F(0)$ given in \eqref{int12}.}
\end{figure}
In Appendix \ref{app1}, we have determined the spectral density $\rho(\tau)$ for the case that $d=2$ or $d=3$ while verifying the $d=4$ result, based on two quite different calculational schemes. Also in the cases $d=2,3$, a positive spectral density is found.

\section{A closer look at the analytic continuation by means of the spectral representation}\label{details}
\subsection{ The case of complex masses}
As already explained in Section \ref{sec:cmpop1}, the knowledge of the spectral density $\rho(\tau)$ allows us to obtain the analytic continuation of the correlation function for complex $k^2$, whereby it not necessarily holds that $k^2>0$. The main formula is given by \eqref{a4}, by means of  which we are able to \emph{define} the analytic continuation of the original momentum space integral \eqref{o1c}, which gives for the correlation function\footnote{We shall work with $d=2$ here for simplicity.},
\begin{equation}\label{h1}
    O(k^2)=\frac{1}{4\pi^2}\int d^2p \frac{1}{\vec{p}^2-i/2}\frac{1}{\vec{p}^2+\vec{k}^2-2\vec{p}\cdot\vec{k}+i/2}\,.
\end{equation}
For any real external momentum vector $\vec{k}$, so that $k^2>0$, the integral \eqref{h1} certainly makes sense. For such vectors, we can reexpress \eqref{h1} as
\begin{equation}\label{ha1}
\frac{1}{4\pi^2}\int d^2p \frac{1}{\vec{p}^2-i/2}\frac{1}{\vec{p}^2+\vec{k}^2-2\vec{p}\cdot\vec{k}+i/2}= \frac{1}{4\pi}\int_0^1 \frac{dx}{xk^2-x^2k^2+ix-i/2}\,,
\end{equation}
as the Feynman (or Schwinger) parametrization trick \eqref{fey-par2} is valid for $k^2>0$. We have again set $2\sqrt{2}\theta^2=1$,\\\\The momentum integral \eqref{h1} is however also well-defined for a more general set of $\vec{k}$. More precisely, even for complex vectors $\vec{k}$, \eqref{h1} can be still well-defined, without having that $F(k^2)$ given by \eqref{ha1} is a valid representation of it, since it is possible that neither the Feynman nor Schwinger trick applies for such vectors $\vec{k}$. In Appendix  \ref{app1}, we have calculated the equivalent of \eqref{int12b} for $d=2$, see eq.~\eqref{new2}. Just as in the previous section, the spectral density $\rho_{d=2}(\tau)$ was obtained using $k^2>0$. It is therefore nontrivial to verify that $O(k^2)=F_1(k^2)$ for any choice of $\vec{k}$ whereby $O(k^2)$ exists, with
\begin{equation}\label{h6}
    F_1(k^2)=\frac{1}{2\pi}\int_1^{\infty}\frac{d\tau}{\tau+k^2}\frac{1}{\sqrt{\tau^2-1}}=\frac{1}{2\pi}\frac{\mathrm{arccos}(k^2)}{\sqrt{1-k^4}}\,,
\end{equation}
which is obtained using the spectral representation \eqref{new2}.\\\\That this constitutes a not so trivial check, becomes more apparent when one realizes that once having obtained \eqref{ha1} for $k^2>0$, one can also use the r.h.s. of \eqref{ha1} to define a function for \emph{any} complex value of $k^2$. Indeed, the integration can be done exactly, by making use of \cite{'tHooft:1978xw}
\begin{equation}\label{g1}
    \int_0^1\frac{dx}{ax+b}=\frac{1}{a}\ln\frac{a+b}{b}\,,
\end{equation}
valid for any complex number $a$ and $b$. The ill-definedness of the integral in the l.h.s. of \eqref{g1} for $\frac{-b}{a}\in[0,1]$ corresponds exactly to the branch cut of the $\ln$ in the r.h.s. of \eqref{g1}. We can rewrite the integral \eqref{ha1} in the following form
\begin{equation}\label{g2}
F_2(k^2)= \frac{1}{4\pi}\int_0^1 \frac{dx}{xk^2-x^2k^2+ix-i/2}=\frac{1}{4\pi}\int_0^1 dx\frac{1}{\sqrt{k^4-1}}\left(\frac{1}{x-x_+}-\frac{1}{x-x_-}\right)
\end{equation}
by introducing partial fractions, with $x_{\pm}=\frac{i+k^2\mp\sqrt{k^4-1}}{2k^2}$. Subsequently, using \eqref{g1} twice, yields
\begin{equation}\label{g3}
F_2(k^2)=  \frac{1}{4\pi}\frac{1}{\sqrt{k^4-1}}\left[\ln\left( i\left(k^2+\sqrt{k^4-1}\right)\right)-\ln \left(i\left(k^2-\sqrt{k^4-1}\right)\right)\right]\,.
\end{equation}
For $k^2>0$, one verifies that $F_1(k^2)=F_2(k^2)$, but the situation changes drastically in the complex $k^2$-plane. For $\mathrm{Re}(k^2)>0$, it stills holds that $F_1(k^2)=F_2(k^2)$, but the functions differ for $\mathrm{Re}(k^2)\leq 0$. This difference can be traced back to the branch cut of $F_2(k^2)$, which is given by the complete imaginary axis. This observation follows immediately as the integral \eqref{g2} is well-defined for any $k^2\in\mathbb{C}$, except for $k^2=is$, $s\in\mathbb{R}$. In sharp contrast, $F_1(k^2)$ displays a cut located on the negative real axis, as it was already explained before.\\\\We shall now argue that only $F_1(k^2)$ obtained via the spectral representation gives a decent analytic continuation of the original momentum integral $O(k^2)$. Let us thus choose an external momentum vector $\vec{k}$ which can be complex, or
\begin{equation}
\vec{k}=\mathrm{Re}(\vec{k})+i\mathrm{Im}(\vec{k})\equiv\vec{k}_R+i\vec{k}_I\,.
\end{equation}
We shall restrict ourselves to complex momenta of the type $\vec{k}=(u+iv,0)$, which allow to reach any value of
\begin{equation}\label{h3}
k^2=\vec{k}^2=u^2-v^2+2iuv\;.
\end{equation}
Although we already chose $k^2$ to be the argument of $O(k^2)$, this is not a priori clear when $\vec{k}$ is not real. However, we  notice that the integral in the r.h.s. of \eqref{h1}, if it at least exists\footnote{We shall shortly see that such $\vec{k}$ do exist.}, will define an analytic function of $k\equiv u+iv$, which can be easily checked by means of the Cauchy-Riemann equations. Since we still have the $O(2)$ rotational symmetry, as we can rotate $\vec{k}_R$ and $\vec{k}_I$ simultaneously, which thereby defines a real rotation of the complex vector $\vec{k}=k\vec{e}_x$, it appears that \eqref{h1} must be a function of $\vec{k}^2$.\\\\We then rewrite \eqref{h1} as
\begin{equation}\label{h2}
    O(k^2)=\frac{1}{4\pi^2}\int dp_x dp_y \frac{1}{p_x^2+p_y^2-i/2}\frac{1}{p_x^2+p_y^2+u^2-v^2+2iuv-2p_x(u+iv)+i/2}\,.
\end{equation}
One could wonder whether it would be possible to obtain finer results if we extend our choice of $\vec{k}$? Namely, using $\vec{k}=(u+iv,u'+iv')$, it is not unconceivable  that one could do better. To  refute this possibility, we notice that we can already choose without loss of generality $\vec{k}=(u+iv,u')$, since we can always rotate our basis to bring $\vec{k}$ into this form. The corresponding integral would read
\begin{equation}\label{h2bis}
    O'(k^2)=\frac{1}{4\pi^2}\int dp_x dp_y \frac{1}{p_x^2+p_y^2-i/2}\frac{1}{p_x^2+p_y^2+u^2-v^2+2iuv+u'^2-2p_x(u+iv)-2p_yu'+i/2}\,,
\end{equation}
but the simple shift $p_y\to p_y+u$ gives $O(k^2)=O'(k^2)$, from which we conclude that we can set $\vec{k}=(u+iv,0)$ without loss of generality.\\\\To proceed, the integral \eqref{h2} is well-defined if the integrand is free of poles for $\vec{p}\in\mathbb{R}^2$. Let us thus check when such poles might appear. The only possibility occurs when\footnote{We can exclude here the $v=0$ case as this corresponds to the anyhow well-defined case of real external momentum.}
\begin{equation}\label{h4}
    p_x=u+\frac{1}{4v}\,,
\end{equation}
which is necessary to kill the imaginary part in the second denominator, and consequently if
\begin{equation}\label{h5}
    \frac{1}{16v^2}-v^2+p_y^2=0\,.
\end{equation}
Hence, if we take $|v|<1/2$, with $u$ arbitrary, the integrand of \eqref{h2} is a regular function of $p_x$ and $p_y$, and the integral will exist as such. It is interesting to notice that this does \emph{not} mean that the Schwinger and/or Feynman trick is applicable for any such vector $\vec{k}$. Nevertheless, the integral has a well-defined value. As such, it must coincide with other ways to compute or define it.\\\\We can now discriminate between $F_1(k^2)$ and $F_2(k^2)$ by taking a test value in the left half plane, such that the original integral \eqref{h2} exists. Its value should then coincide with either $F_1(k^2)$ or $F_2(k^2)$. Actually, we can immediately motivate that $F_2(k^2)$ cannot be correct in the whole complex plane. Looking at \eqref{h3}, it is clear that $k^2\in i\mathbb{R}$ if $u=\pm v$, in which case we have $k^2=\pm2iv^2$. But as we have shown, for $|v|<1/2$, the original integral is well-defined, meaning that there cannot be a cut on the \emph{whole} imaginary axis, whereas $F_2(k^2)$ has a cut for any $k^2\in i\mathbb{R}$. This shows that $F_2(k^2)$ cannot be the analytic continuation of the original momentum integral. In fact, $F_1(z)$ also describes the analytical continuation of $F_2(z)$ with $\textrm{Re}(z)>0$ to the left half complex plane, defined by $\textrm{Im}(z)\leq0$. Since the whole imaginary axis is a cut of $F_2(z)$, it is impossible to find an open region of $\mathbb{C}$ where $F_2(z)_{\textrm{Im}(z)>0}$ and $F_2(z_2)_{\textrm{Im}(z)<0}$, coincide. Said otherwise, $F_2(z)_{\textrm{Im}(z)>0}$ and $F_2(z_2)_{\textrm{Im}(z)<0}$ are not analytically connected.\\\\As a second illustration, let us look at the test value
\begin{equation}\label{h8}
    (u_*,v_*)=\left(\frac{1}{4},\frac{1}{3}\right)\Rightarrow k_*^2=\frac{-7}{144}+\frac{i}{6}\,.
\end{equation}
It can be checked that in this case, we have
\begin{equation}\label{h9}
    F_1(k_*^2)\approx 0.254-0.028i,\qquad F_2(k_*^2)\approx -0.239-0.024i\,.
\end{equation}
We can also calculate \eqref{h2} directly. We write
\begin{equation}\label{h10a}
   O(k_*^2)=\frac{1}{4\pi^2}\int_{-\infty}^{+\infty} dp_y \left[\int_{-\infty}^{+\infty} dp_x\frac{1}{p_x^2+p_y^2-i/2}\frac{1}{p_x^2+p_y^2+u_*^2-v_*^2+2iu_*v_*-2p_x(u_*+iv_*)+i/2}\right]\,,
\end{equation}
and we use the residue theorem for the $p_x$-integration. We close the contour along the upper hemisphere at $\infty$, where the function vanishes, and we notice that 2 of the 4 poles are located within the contour, being
\begin{equation}\label{h11}
    p_{x1}=\frac{\sqrt{i-2p_y^2}}{\sqrt{2}}\,,\qquad p_{x2}=\frac{1}{12}\left(3+4i-6\sqrt{2}\sqrt{-i-2p_y^2}\right)\,,
\end{equation}
which is independent of $p_y$. So,
\begin{eqnarray}\label{h12}
     O(k_*^2)&=&\frac{36\sqrt{2}}{\pi}i\int_{-\infty}^{+\infty} dp_y f(p_y)\nonumber\\
    f(p_y)&=&\frac{1}{\sqrt{-i-2p_y^2}\left(7+120i+(36+48i)\sqrt{-2i-4p_y^2}\right)}-\frac{1}{\sqrt{i-2p_y^2}\left(7-168i+(36+48i)\sqrt{2i-4p_y^2}\right)}\,.\nonumber\\
\end{eqnarray}
Since $p_y\in\mathbb{R}$, the remaining integral is well-defined\footnote{We will never come into the vicinity of the cut of the occurring roots.} and can be computed using direct primitivation. After the smoke clears, one finds\footnote{We did not write the closed expression, as it is rather intransparent.}
\begin{eqnarray}\label{h13}
    O(k_*^2)\approx 0.254-0.028i\,,
\end{eqnarray}
a value which nicely coincides with $F_1(k_*^2)$, while differing from $F_2(k_*^2)$.\\\\We have provided  a nontrivial verification of the correctness of the analytically continued function $F_1(k^2)$. In fact, much more can be learned from our analysis. Let us take expression \eqref{h3} into reconsideration,
\begin{equation}\label{h3again}
k^2=u^2-v^2+2iuv\,.
\end{equation}
As we have shown, problems can only occur for  $|v|\geq1/2$. But even restricting ourselves to $|v|<1/2$, eq.~\eqref{h3again} means that we can still reach a relatively large region of the whole complex $k^2$-plane without encountering difficulties, since $u$ is arbitrary. For example, for $u=0$, we can end up on the negative real axis, but there can be a cut on the negative real axis only for  $k^2\leq-1/4$.   As another example, we repeat that by taking $u=\pm v$, we conclude that a cut on the imaginary axis is a priori possible, but only when  $|\mathrm{Im}(k^2)|\geq 1/2$.\\\\In FIG.~2, the shaded parabolic region corresponds to the values of $k^2\in\mathbb{C}$  for which the integral \eqref{h2} is certainly well-defined. We clearly observe a small region in the left half plane where $F(k^2)$ should be well-defined, even without invoking analytical continuation.
 \begin{figure}[h]
  \begin{center}
\includegraphics[width=7cm]{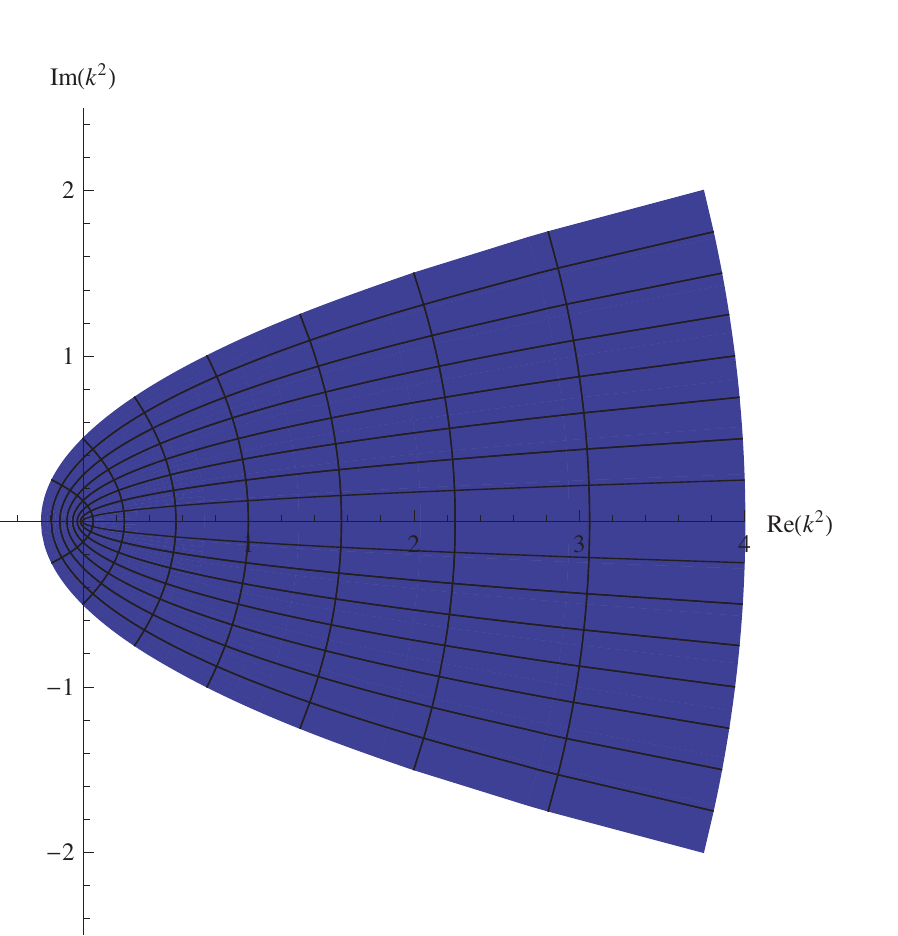}\label{cone}
     \end{center}
  \caption{Parabolic region in the complex $k^2$ plane  where the original momentum integral $O(k^2)$ exists. }
\end{figure}
The results obtained do, of course, not mean that there \emph{must} be a cut for $k^2\in[-\infty,-1/4]$ or that there is a ``cut region'' outside of the displayed parabola. But we have discovered a certain region where the analytic continuation has to be consistent with the value of the integral. Certainly, the analytic continuation of $F_1(k^2)$ is consistent with the value of the integral within the boundaries of FIG.~2. \\\\To close this section, we mention that we could also have started with the alternative definition
\begin{equation}\label{h14}
    \hat{O}(k^2)=\frac{1}{4\pi^2}\int d^2p \frac{1}{\vec{p}^2+\vec{k}^2-2\vec{p}\cdot\vec{k}-i/2}\frac{1}{\vec{p}^2+i/2}\,,
\end{equation}
which corresponds to switching the momenta running in the two legs. For real external momenta $\vec{k}$, we obviously have $O(k^2)=\hat{O}(k^2)$, but for complex $\vec{k}$, we can no longer perform a translation on the real integration momentum $\vec{p}$ to prove this. \\\\However, again setting $\vec{k}=(u+iv,0)$, it is easily checked that $|v|<1/2$ is sufficient to also guarantee that $\hat{O}(k^2)$ is well-defined, while from FIG.~\ref{fig3a}, \ref{fig3b}, \ref{fig4a}, \ref{fig4b}, it is clear that  $O(k^2)=\hat{O}(k^2)$ for $u$ arbitrary, $|v|<1/2$. For completeness, we have also shown
$F_1(k^2)$ in FIG.~\ref{fig3c}, \ref{fig4c}, which illustrates that indeed $O(k^2)=\hat{O}(k^2)=F_1(k^2)$ over the parabolic $k^2$-region shown in FIG.~2. To avoid confusion, we point out that we have plotted $O(u,v)\equiv O(k^2)$, with $k^2=u^2-v^2+2iuv$, and analogously for the other functions. We notice that this is equivalent with having the Schwartz reflection principle, $O(k^2{}^\dagger)=O(k^2){}^\dagger$, a property shared with $F_1(k^2)$.\\

\begin{figure}[h]
  \begin{center}
    \subfigure[]{\includegraphics[width=4.8cm]{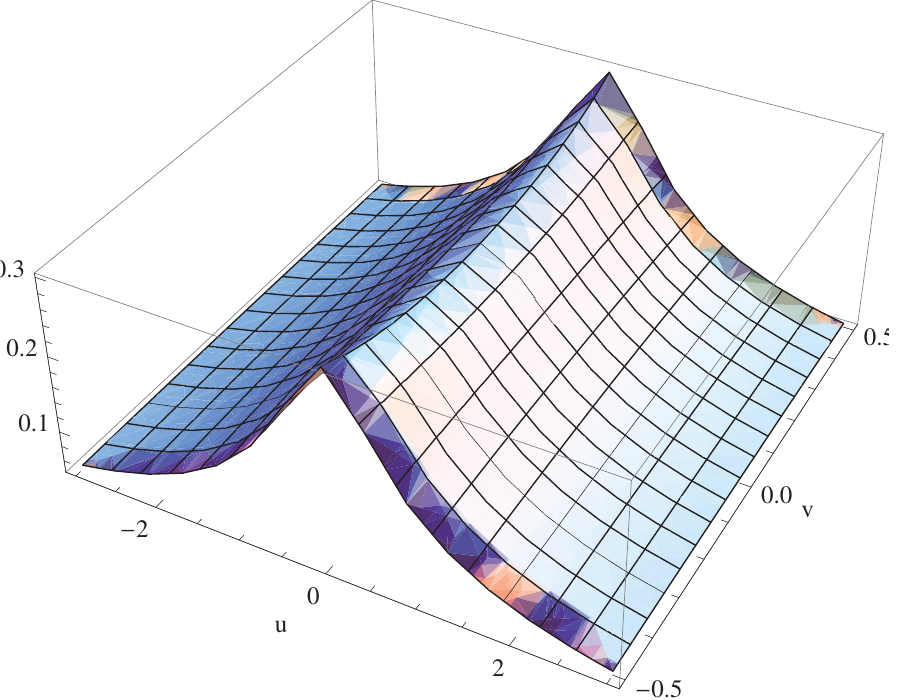} \label{fig3a}}
    \hspace{0.5cm}
    \subfigure[]{\includegraphics[width=4.8cm]{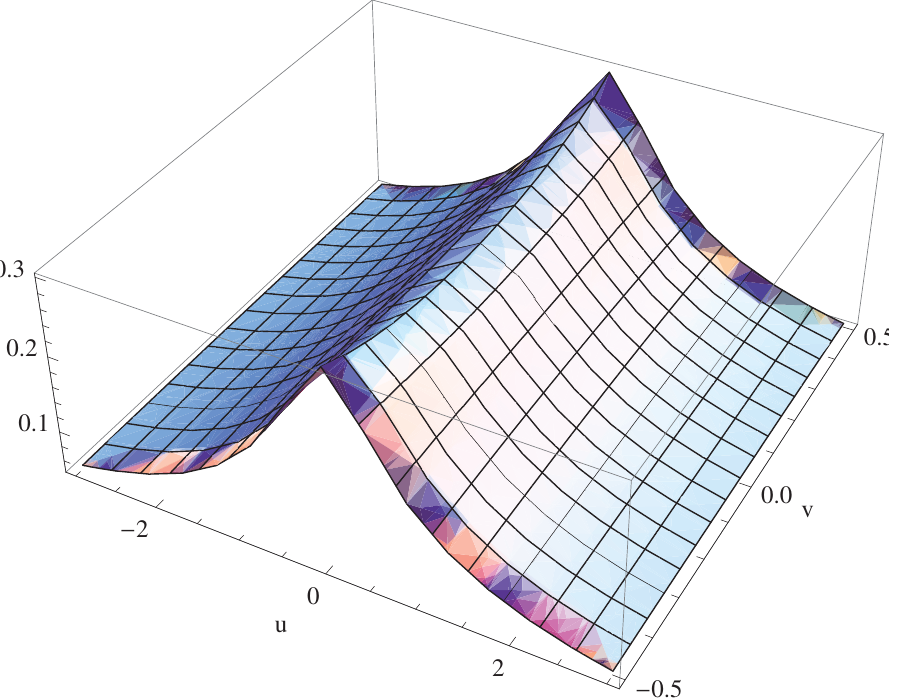}\label{fig3b}}
    \hspace{0.5cm}
        \subfigure[]{\includegraphics[width=4.8cm]{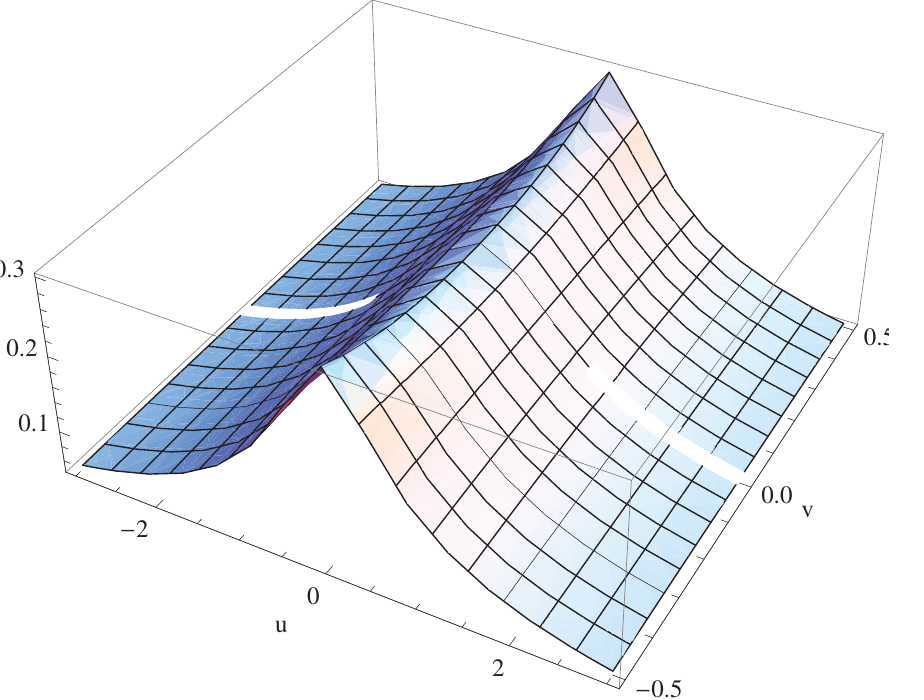}\label{fig3c}}
     \end{center}
  \caption{$\mathrm{Re}(O)$, $\mathrm{Re}(\hat{O})$ and $\mathrm{Re}(F_1)$. }
\end{figure}

\begin{figure}[h]
  \begin{center}
    \subfigure[]{\includegraphics[width=4.8cm]{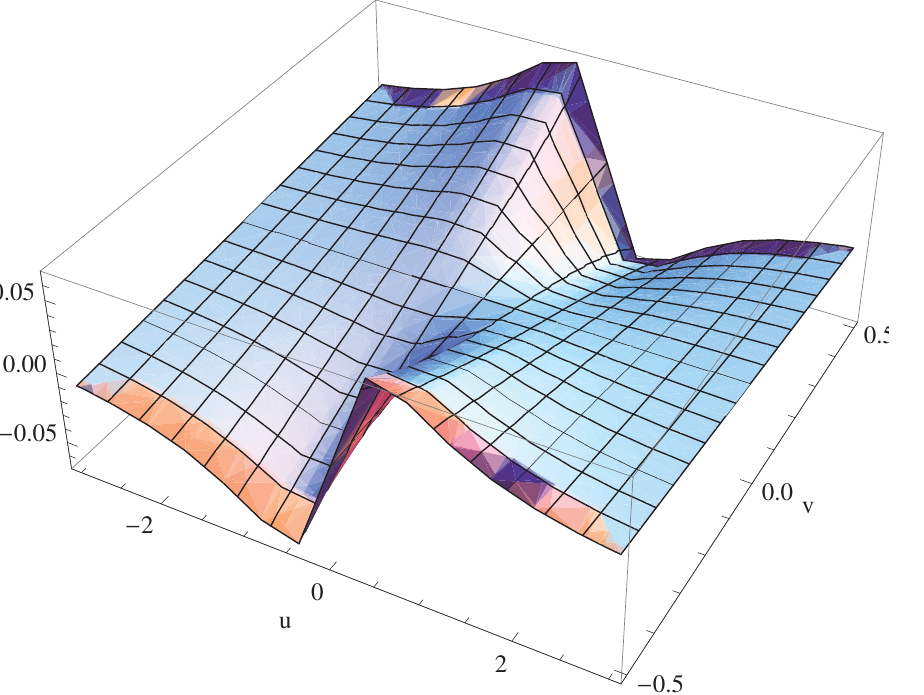} \label{fig4a}}
    \hspace{0.5cm}
    \subfigure[]{\includegraphics[width=4.8cm]{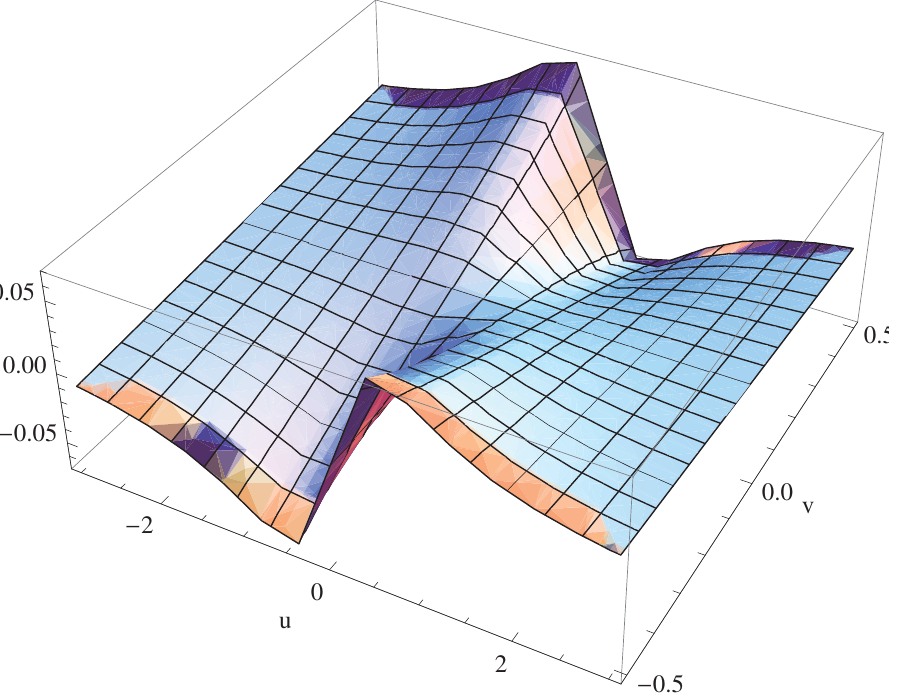}\label{fig4b}}
    \hspace{0.5cm}
        \subfigure[]{\includegraphics[width=4.8cm]{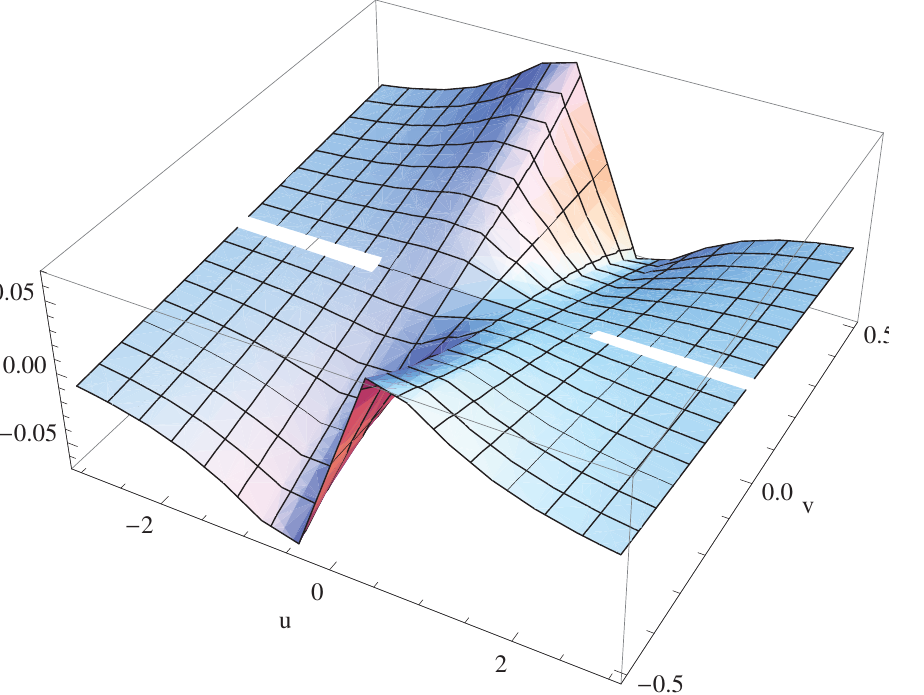}\label{fig4c}}
     \end{center}
  \caption{$\mathrm{Im}(O)$, $\mathrm{Im}(\hat{O})$ and $\mathrm{Im}(F_1)$. }
\end{figure}
Although we focused exclusively on the $d=2$ case in this section, similar conclusions can be drawn for $d=3$ or $d=4$.\\

\subsection{The case of two real masses}\label{detailsb}
In order to corroborate the previous nontrivial conclusions about a relatively simple Feynman integral with complex masses, we find it instructive to also include a similar analysis of the probably more familiar case of two real and positive masses, being $m_1$ and $m_2$. This has been investigated in great detail in e.g.~\cite{Itzykson:1980rh}, albeit in Minkowskian space. The analog of \eqref{h1} is given by the following correlation function,
\begin{equation}\label{h10}
    O_2(k^2)=\frac{1}{4\pi^2}\int d^2p \frac{1}{\vec{p}^2+ m_1^2}\frac{1}{\vec{p}^2+\vec{k}^2-2\vec{p}\cdot\vec{k}+ m_2^2 }\,.
\end{equation}
Translating the results of \cite{Itzykson:1980rh} to Euclidean space, it was shown that this function of $k^2$ has a positive spectral density in combination with a branch cut on the negative real $k^2$-axis starting from $-(m_1+ m_2)^2$ until $-\infty$.\\\\Let us now also investigate the region where this integral actually exists. We can again restrict ourselves to complex momenta of the type $\vec{k} = (u + iv,0)$ without loss of generality. Inserting this in eq.~\eqref{h10}, we obtain,
\begin{equation}\label{h2a}
    O_2(k^2)=\frac{1}{4\pi^2}\int dp_x dp_y \frac{1}{p_x^2+p_y^2+ m_1^2}\frac{1}{p_x^2+p_y^2+u^2-v^2+2iuv-2p_x(u+iv)+m_2^2}\,.
\end{equation}
Let us check when poles can emerge. The first denominator is always positive, however, the second denominator can have poles when the imaginary part vanishes,
\begin{eqnarray}\label{im}
2 i u v - 2 i p_x v = 0 \Leftrightarrow  v = 0 \quad\text{   or  }\quad u = p_x\;,
\end{eqnarray}
simultaneously with the real part being equal to zero,
\begin{eqnarray}
p_x^2 + p_y^2 + u^2 - v^2 + 2 p_x u + m_2^2\;.
\end{eqnarray}
Inserting the first solution of eq.~\eqref{im},\textit{i.e.}~$v=0$, in the equation above, we find
\begin{eqnarray}
(p_x + u)^2 + p_y^2 = - m_2^2\;,
\end{eqnarray}
which can never be fulfilled. The second solution, \textit{i.e.}~$u=p_x$, results in
\begin{eqnarray}
 p_y^2  &=& v^2 - m_2^2\;.
\end{eqnarray}
Therefore, no poles shall occur for $|v| < m_2$, while for $|v|\geq m_2$, the integral \eqref{h10} becomes ill-defined.
This is important as it shows us that also in the well-studied case of two real masses, the original integral \eqref{h10} is also only well-defined in a certain region of the complex plane, here displayed in FIG.~5. Consequently, one also needs to perform an analytic continuation outside this region, just as in the case of pure complex masses.
\begin{figure}[h]
  \begin{center}
\includegraphics[width=7cm]{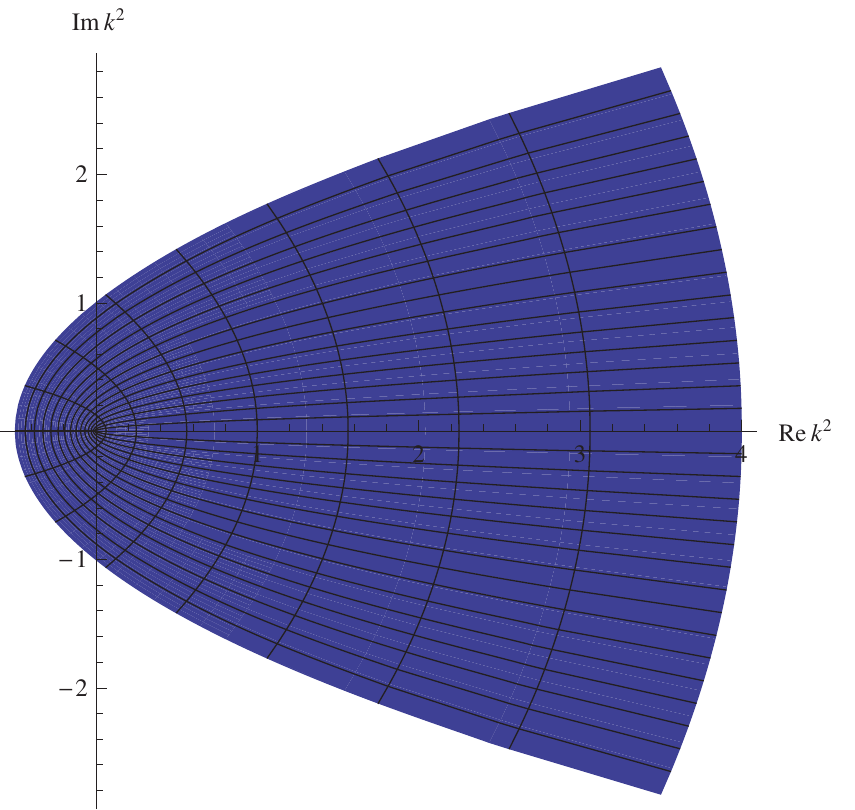}\label{cone2}
     \end{center}
  \caption{Parabolic region in the complex $k^2$ plane  where the momentum integral $O_2(k^2)$ exists with $m_2^2 = 1/2$.}
\end{figure}

\section{Higher loop case}\label{higherloop}
Having discussed in detail the analyticity properties of one loop correlation functions made of $i$-particles, we face now the extension to higher orders. As we shall see, the use of $i$-particles turns out to be extremely useful here, enabling us to construct a rather interesting iterative procedure valid for an $n$-loop integral. Note that in the toy model only one type of diagrams appear, sometimes known as water melon diagrams, because there are no interactions. The case of diagrams including interactions as appearing for the GZ theory is left for future studies. Let us begin with the two loop order.
\subsection{A two loop example}
In order to obtain an example of a  two loop correlation function, we consider the local operator
\begin{align}
O_2(x) = \lambda(x) \eta(x) U(x) \;. \label{o2}
\end{align}
For the correlation function $\langle O_2(k) O_2(-k) \rangle$  we find
\begin{align}
\langle O_2(k) O_2(-k) \rangle  = J_2(k^2)= \int {\frac{d^4p}{(2\pi)^2}}  {\frac{d^4q}{(2\pi)^2}} \frac{1}{q^2} \frac{1}{p^2+i\sqrt{2}\theta^2} \frac{1}{(k-q-p)^2-i\sqrt{2}\theta^2} \;, \label{a1}
\end{align}
where $k^2$ is the Euclidean external momentum. In order to evaluate it\footnote{From now on, we shall not bother about potential subtractions to make it well-defined. These can always be obtained by taking a suitable number of derivatives w.r.t.~the external momentum $k^2$.}, we rewrite it as
\begin{align}
J_2(k^2) = \int {\frac{d^4q}{(2\pi)^4}} \frac{1}{q^2} \left( \int {\frac{d^4p}{(2\pi)^2}}   \frac{1}{p^2+i\sqrt{2}\theta^2} \frac{1}{(k-q-p)^2-i\sqrt{2}\theta^2} \right) \;, \label{a2}
\end{align}
so that we can use the spectral representation \eqref{int11b} already found at one loop level, namely
\begin{align}
\int {\frac{d^4p}{(2\pi)^2}}   \frac{1}{p^2+i\sqrt{2}\theta^2} \frac{1}{(k-q-p)^2-i\sqrt{2}\theta^2} = \int_{\tau_{0}}^{\infty} d\tau \; \rho({\tau}) \frac{1}{\tau+ (k-q)^2} \;, \label{a3}
\end{align}
with $\tau_0=2\sqrt{2}\theta^2$. Therefore
\begin{align}
J_2(k^2) = \int_{\tau_{0}}^{\infty} d\tau \; \rho({\tau}) \left( \int {\frac{d^4q}{(2\pi)^2}} \frac{1}{q^2} \frac{1}{\tau+ (k-q)^2}\right) \,.
\end{align}
The $q$-integral becomes now straightforward. It corresponds to the one loop integral in which one particle is massless, and the other one has a real and positive mass $\tau$. Such integrals also have a spectral representation with positive spectral density, given by
\begin{equation}
\int {\frac{d^4q}{(2\pi)^4}} \frac{1}{q^2} \frac{1}{\tau+ (k-q)^2} =  \int_{\tau}^{\infty}  ds\; \rho_{1}(s) \frac{1}{s+k^2}  \,,
\end{equation}
where $\rho_1(s)$ is the associated spectral function. Its explicit value can be found in Appendix \ref{app1}, eq.~\eqref{appspec}. Thus,
\begin{align}\label{j2}
J_2(k^2) = \int_{\tau_{0}}^{\infty} d\tau \; \rho({\tau}) \int_{\tau}^{\infty}  ds \;\rho_{1}(s) \frac{1}{s+k^2}\,.
\end{align}
This is not yet in the form of a spectral representation. By switching the order of integration, we can however reexpress the double integral as
\begin{equation}\label{dd1}
    J_2(k^2)= \int_{\tau_0}^\infty ds \frac{1}{s+k^2}\int_{\tau_0}^{s}d\tau\rho_1(s)\rho(\tau)=\int_{\tau_0}^{\infty}\frac{\rho_2(s)}{s+k^2}ds\,.
\end{equation}
This means that the spectral density is given by
\begin{equation}\label{dd2}
    \rho_2(s)= \int_{\tau_0}^{s}d\tau\rho(\tau)\rho_1(s)\,.
\end{equation}
whereby we kept in mind that $\rho_1(s)$ will depend on $\tau$ as well. This allows to make the following observations:
\begin{itemize}
\item The branch cut will be located on the negative real axis from $-\infty$ to $-\tau_0$.
\item For $s\in[\tau_0,\infty]$, we have $\rho(\tau)\geq 0$ due to the positivity of $\rho(\tau)$ over the interval $[\tau_0,\infty]$. We shall thus have that $\rho_2(s)\geq 0$ itself if $\rho_1(s)\geq 0$ for $s\in[\tau_0,\infty]$. The latter turns out to be true, as $\rho_1(s)$ is defined for all $\tau$ with $\tau\geq \tau_0$, where it is positive.
\end{itemize}
In conclusion, the two loop correlation function \eqref{a1} can also be reexpressed in a spectral form, with associated positive spectral density. As we have seen, there is even no need to explicitly evaluate the integral or spectral density to establish these conclusions. To completely convince ourselves, we can also evaluate \eqref{dd2} explicitly, by making use of \eqref{int12b} and \eqref{appspec}, which leads to
\begin{eqnarray}\label{dd3}
    \rho_2(s)&=&\frac{1}{(16 \pi^2)^2} \int_{2 \sqrt{2} \theta^2 }^{\infty} d \tau \frac{\sqrt{\tau^2 - 8 \theta^4}}{\tau}   \left( 1- \frac{\tau}{ s} \right) \nonumber\\
    &=& \frac{1}{(16 \pi^2)^2}  \left(\frac{1}{2} \sqrt{s^2- 8 \theta ^4} + 2  \sqrt{2}\theta ^2 \left(\arctan \left[\frac{2 \sqrt{2} \theta ^2}{\sqrt{s^2-8 \theta ^4}}\right] - \frac{\pi}{2}  \right) +  \frac{4 \theta^4}{ s} \left(  \ \ln \frac{s+\sqrt{s^2-8 \theta ^4}}{2 \sqrt{2} \theta^2} \right) \right)\,,
\end{eqnarray}
and then one can check that $\rho_2(s)$ is indeed positive for $s\geq 2\sqrt{2}\theta^2$.

\subsection{A three loop example}
To construct another example of a higher loop correlation function, one might consider the composite operator
\begin{align}
O_3(x) = \frac{1}{4} {\left( \lambda(x) \eta(x) \right) }^2 \;, \label{o3}
\end{align}
which gives rise to a three loop expression
\begin{align}
J_3(k^2) = \int \frac{d^dq}{(2\pi)^d}\frac{d^d\ell}{(2\pi)^d}\frac{d^dp}{(2\pi)^d}
	\frac{1}{q^2+i \sqrt{2}\theta^2}\frac{1}{\ell^2-i  \sqrt{2}\theta^2}\frac{1}{p^2+i \sqrt{2}\theta^2}\frac{1}{(k-q-\ell-p)^2-i \sqrt{2} \theta^2}\,.
\end{align}
We proceed in a similar way as for the two loop case. Starting from
\begin{align}
\int \frac{d^dp}{(2\pi)^d} \frac{1}{p^2+i \sqrt{2}\theta^2}\frac{1}{(k-q-\ell-p)^2-i  \sqrt{2}\theta^2} =  \int_{\tau_{0}}^{\infty} d\tau \; \rho({\tau}) \frac{1}{\tau+(k-p-\ell)^2} \;, \label{b2}
\end{align}
and by making the change of variables
\begin{equation}
\ell \rightarrow \ell-q  \;, \label{d3}
\end{equation}
we get
\begin{align}
J_3(k^2) =  \int_{\tau_{0}}^{\infty} d\tau \; \rho({\tau}) \int \frac{d^d\ell}{(2\pi)^d} \frac{1}{\tau+(k-\ell)^2}  \int \frac{d^dq}{(2\pi)^d} \frac{1}{q^2+i\sqrt{2}\theta^2} \frac{1}{(\ell-q)^2-i\sqrt{2}\theta^2} \;.  \label{d4}
\end{align}
Thus,
\begin{align}
J_3(k^2) =   \int_{\tau_{0}}^{\infty} d\tau \; \rho({\tau}) \int \frac{d^d\ell}{(2\pi)^d} \frac{1}{\tau+(k-\ell)^2}
 \int_{\tau_{0}}^{\infty} ds\; \rho(s) \frac{1}{s+\ell^2}   \;,  \label{d5}
\end{align}
from which
\begin{align}
J_3(k^2) = \int_{\tau_0}^\infty d\tau \rho(\tau)\underline{\int_{\tau_0}^{\infty}ds \rho(s)\int_{(\sqrt{\tau}+\sqrt{s})^2}^{\infty}dr\rho_1(r)\frac{1}{r+k^2}} \label{d1}
\end{align}
follows, keeping in mind that the branch cut for two particles with respective positive real masses $\mu_1^2$ and $\mu_2^2$ starts from $-(\mu_1+\mu_2)^2$, see also Section \ref{detailsb}.\\\\We see that we again find a kind of convolution of already known spectral representations. As before, we still need to show that $J_3(k^2)$ can be brought into a spectral representation with positive density $\rho_3(r)$ for $r\geq \tau_*\geq 0$, thereby establishing a physical branch cut along $[-\infty,-\tau_*]$, with $\tau_*$ yet unknown. It turns out that this is little more complicated than in the two loop case. Let us look at the underlined part of \eqref{d1} first, which is a generalization of the previous case $J_2(k^2)$. Hence, we wish to examine
\begin{equation}\label{dd5}
    \overline{J}_2(k^2)= \int_{\tau_0}^{\infty}ds \rho(s)\int_{(\sqrt{s}+a)^2}^{\infty}dr\rho_1(r)\frac{1}{r+k^2}\,,
\end{equation}
where $a\geq \sqrt{\tau_0}$. We propose the substitution $\sqrt{\xi}=\sqrt{s}+a$, leading to
\begin{equation}\label{dd6}
    \overline{J}_2(k^2)= \int_{(\sqrt{\tau_0}+a)^2}^{\infty}d\xi \overline{\rho}(\xi)\int_{\xi}^{\infty}dr\rho_1(r)\frac{1}{r+k^2}\,,
\end{equation}
where we have introduced a novel spectral density
\begin{equation}\label{dd7}
    \overline{\rho}(\xi)= \frac{\sqrt{\xi}-a}{\sqrt{\xi}}\rho\left[\left(\sqrt{\xi}-a\right)^2\right]\,.
\end{equation}
It is clear that for $\xi\geq(\sqrt{\tau_0}+a)^2$, we have $\overline{\rho}(\xi)\geq 0$, invoking the properties of $\rho(s)$. We then observe that we have managed to reduce $\overline{J}_2(k^2)$ to the $J_2(k^2)$ case already studied, see eq.~\eqref{j2}, meaning that \eqref{dd5} has a spectral density representation of the form
\begin{equation}\label{8}
    \overline{J}_2(k^2)=\int_{(\sqrt{\tau_0}+a)^2}^{\infty}ds\overline{\rho}_2(s)\frac{1}{s+k^2},\qquad \textrm{with}\;\overline{\rho}_2(s)\geq0\,.
\end{equation}
As such, we can write
\begin{equation}\label{9}
    J_3(k^2)= \int_{\tau_0}^\infty d\tau \rho(\tau)\int_{(\sqrt{\tau}+\sqrt{\tau_0})^2}^{\infty}ds\overline{\rho}_2(s)\frac{1}{s+k^2}\,.
\end{equation}
This is again of the type $\overline{J}_2(k^2)$, from which we conclude that we can write
\begin{equation}\label{10}
    J_3(k^2)= \int_{(\sqrt{\tau_0}+\sqrt{\tau_0})^2}^{\infty}ds \rho_3(s)\frac{1}{s+k^2}=\int_{4\tau_0}^{\infty}ds \rho_3(s)\frac{1}{s+k^2}\,,
\end{equation}
with $\rho_3(s)\geq0$, for $s\geq \tau_*\equiv8\sqrt{2}\theta^2$.\\\\We may summarize by noticing that the higher loop correlation functions can be written as a kind of convolution of the lower loop spectral representations. If the involved lowest, \textit{i.e.}~ two loop, spectral densities are physical, so will the higher loop spectral densities be, by making use of the iterative argument given in this section.

\section{A few remarks on the physical meaning of the $i$-fields $(\lambda,\eta)$} \label{i-fields}
As we have seen in the previous sections, the introduction of the $i$-fields $(\lambda,\eta)$ enables us to construct in a relatively simple way a set of composite operators whose correlation functions exhibit real cuts only. It is worth thus spending a few words on the physical meaning of these fields. \\\\Looking at the very starting point, eq.~\eqref{act}, we see that the action of the field $\psi$ has been modified by the addition of a nonlocal term, accounting for a deep modification of the long range behavior of the model. It is easy to figure out that the field $\psi$ itself acquires thus a long range nonlocal component. The construction of a suitable set of composite operators in terms of the field $\psi$ becomes a highly nontrivial task, as nonlocal terms have to be incorporated in order to achieve meaningful correlation functions. When the model is cast in local form, the long range behavior of the theory is accounted for by the auxiliary localizing fields $\vp, \bar \vp$.  The fields $\psi, \vp, {\bar \vp}$ have to be considered thus on an equal footing. It is instructive now to express the composite operator $O_1(x)=\lambda(x) \eta(x)$ in terms of the original field $\psi$.
From the equations of motion, it follows
\begin{align}
\vp & =\frac{\theta^2}{(-\partial^2)} \psi \;, \nonumber \\
{\bar \vp} & =- \frac{\theta^2}{(-\partial^2)} \psi \;. \label{eqsm}
\end{align}
Analogously, for the $i$-fields, one gets
\begin{align}
\lambda & =\frac{1}{\sqrt{2}} \left( \psi -i\sqrt{2} \frac{\theta^2}{(-\partial^2)} \psi \right)\;, \nonumber \\
\eta & =\frac{1}{\sqrt{2}} \left( \psi +i\sqrt{2} \frac{\theta^2}{(-\partial^2)} \psi \right)\;. \label{etal}
\end{align}
Therefore, for the operator $O_1(x)$  one obtains
\begin{equation}
O_1(x) = \lambda(x)\eta(x)= \frac{1}{2} \left( \psi^2 + 2 \left( \frac{\theta^2}{(-\partial^2)} \psi \right)^2 \right) \;, \label{nlo1}
\end{equation}
from which one clearly sees that nonlocal terms are needed to achieve a sensible operator. The $i$-fields  $\lambda, \eta$ provide thus the correct field variables which enable us to construct in a simple way,  and within a local quantum field theory framework,  the relevant composite operators of the theory  by taking into account the nonlocal long range effects.

\section{Introducing $i$-particles for the Gribov-Zwanziger action} \label{gz}
Let us discuss here how $i$-particles can arise in the Gribov-Zwanziger action, given by expression (\ref{locact1}). In what follows, we shall limit  ourselves to evaluate correlation functions of suitable composite operators at one loop order only. To introduce the $i$-particles in the Gribov-Zwanziger action, it suffices thus to consider the quadratic part of expression (\ref{locact1})  containing only the fields $(A^a_{\mu}, \vp_\mu^{ab}, {\bar\vp}_\mu^{ab})$, namely
\begin{align}
S_{GZ}^{\rm quad} = \int d^4x\; \left( \frac{1}{2} A^a_{\mu} (-\partial^2) A^a_{\mu} + {\bar \vp}^{ab}_{\mu}(-\partial^2)\vp^{ab}_{\mu} +\gamma^2\,g\,f^{abc}A_\mu^{a}(\vp_\mu^{bc}-{\bar\vp}_\mu^{bc}) \right) \;, \label{quad}
\end{align}
where use has been made of the transversality of the gauge field, $\partial_{\mu}A^a_{\mu}=0$. We now proceed by  decomposing the fields $(\vp_\mu^{ab}, {\bar\vp}_\mu^{ab})$ in symmetric and anti-symmetric components in color space, {\it i.e.}
\begin{align}
\vp^{ab}_{\mu} & = \vp^{[ab]}_{\mu} + \vp^{(ab)}_{\mu}  \;,  \nonumber \\
\vp^{[ab]}_{\mu}  & = \frac{1}{2} \left( \vp^{ab}_{\mu} - \vp^{ba}_{\mu} \right) \;, \nonumber \\
\vp^{(ab)}_{\mu}  & = \frac{1}{2} \left( \vp^{ab}_{\mu} + \vp^{ba}_{\mu} \right) \;, \label{svp}
\end{align}
and
\begin{align}
{\bar \vp}^{ab}_{\mu} & = {\bar \vp}^{[ab]}_{\mu} + {\bar \vp}^{(ab)}_{\mu}  \;, \nonumber \\
{\bar \vp}^{[ab]}_{\mu}  & = \frac{1}{2} \left( {\bar \vp}^{ab}_{\mu} - {\bar \vp}^{ba}_{\mu} \right) \;, \nonumber \\
{\bar \vp}^{(ab)}_{\mu}  & = \frac{1}{2} \left( {\bar \vp}^{ab}_{\mu} + {\bar \vp}^{ba}_{\mu} \right) \;. \label{bsvp}
\end{align}
Thus
\begin{align}
S_{GZ}^{\rm quad} = \int d^4x\; \left( \frac{1}{2} A^a_{\mu} (-\partial^2) A^a_{\mu} + {\bar \vp}^{[ab]}_{\mu}(-\partial^2)\vp^{[ab]}_{\mu} + {\bar \vp}^{(ab)}_{\mu}(-\partial^2)\vp^{(ab)}_{\mu} +\gamma^2\,g\,f^{abc}A_\mu^{a}(\vp_\mu^{[bc]}-{\bar\vp}_\mu^{[bc]}) \right) \;. \label{quad1}
\end{align}
A first  step towards diagonalization of this expression is achieved by setting
\begin{align}
\vp^{[ab]}_{\mu} = \frac{1}{\sqrt{2}} \left( U^{[ab]}_{\mu} + i V^{[ab]}_{\mu} \right) \;, \nonumber \\
{\bar \vp}^{[ab]}_{\mu} = \frac{1}{\sqrt{2}} \left( U^{[ab]}_{\mu} - i V^{[ab]}_{\mu} \right)  \;, \label{vu}
\end{align}
so that
\begin{align}
S_{GZ}^{\rm quad} = \int d^4x\; \left( \frac{1}{2} A^a_{\mu} (-\partial^2) A^a_{\mu} + \frac{1}{2} V^{[ab]}_{\mu}(-\partial^2)V^{[ab]}_{\mu} + i\sqrt{2} g\gamma^2 f^{abc}A^a_{\mu} V^{[bc]}_{\mu}
+  \frac{1}{2} U^{[ab]}_{\mu}(-\partial^2)U^{[ab]}_{\mu} +  {\bar \vp}^{(ab)}_{\mu}(-\partial^2)\vp^{(ab)}_{\mu} \right) \;. \label{quad2}
\end{align}
From expression (\ref{quad2}) one sees that the gauge field $A^a_{\mu}$ mixes with the adjoint projection of $V^{[ab]}_{\mu}$, obtained by employing the following decomposition
\begin{align}
V^{[ab]}_{\mu}= \frac{1}{N} f^{abp}f^{pmn} V^{[mn]}_{\mu} + \left( V^{[ab]}_{\mu} - \frac{1}{N} f^{abp}f^{pmn} V^{[mn]}_{\mu} \right) = f^{abp} V^{p}_{\mu} + S^{[ab]}_{\mu} \;, \label{dec}
\end{align}
where
\begin{align}
V^{p}_{\mu}= \frac{1}{N} f^{pmn} V^{[mn]}_\mu \;, \label{adjp}
\end{align}
stands for the adjoint projection of $V^{[ab]}_{\mu}$ in color space, and
\begin{align}
S^{[ab]}_{\mu} = V^{[ab]}_{\mu} - \frac{1}{N} f^{abp}f^{pmn} V^{[mn]}_{\mu} \;, \label{indep}
\end{align}
denote the remaining independent components of $V^{[ab]}_{\mu}$ which are orthogonal to the tensor $f^{abc}$. In fact, making use of
\begin{align}
f^{abc} f^{dbc} = N \delta^{ad} \;, \label{ff1}
\end{align}
it is easily checked that
\begin{align}
f^{abc} S^{[ab]}_{\mu}=0 \;. \label{ort}
\end{align}
 Therefore, expression (\ref{quad2}) becomes
\begin{align}
S_{GZ}^{\rm quad}  =  \int d^4x  & \left( \frac{1}{2} A^a_{\mu} (-\partial^2) A^a_{\mu} + \frac{N}{2} V^{a}_{\mu}(-\partial^2) V^{a}_{\mu} + i\sqrt{2} g \gamma^2 N A^a_\mu V^a_\mu  \right) \nonumber \\
+ \int d^4x & \left( \frac{1}{2} S^{[ab]}_{\mu} (-\partial^2) S^{[ab]}_{\mu} +  \frac{1}{2} U^{[ab]}_{\mu}(-\partial^2)U^{[ab]}_{\mu} +  {\bar \vp}^{(ab)}_{\mu}(-\partial^2)\vp^{(ab)}_{\mu} \right) \;. \label{quad3}
\end{align}
Finally, setting
\begin{align}
A^{a}_{\mu} & = \frac{1}{\sqrt{2}} \left( \lambda^{a}_{\mu}+ \eta^{a}_{\mu} \right) \;, \nonumber \\
{V}^{a}_{\mu} & = \frac{1}{\sqrt{2N}} \left( \lambda^{a}_{\mu}- \eta^{a}_{\mu} \right) \;, \label{fv}
\end{align}
one obtains the diagonal action
\begin{align}
S_{GZ}^{\rm quad} = \int d^4x & \left( \frac{1}{2} {\lambda}^{a}_{\mu} \left(-\partial^2+ i\sqrt{2N}g\gamma^2 \right) {\lambda}^{a}_{\mu} + \frac{1}{2} {\eta}^{a}_{\mu}\left(-\partial^2- i\sqrt{2N}g\gamma^2\right){\eta}^{a}_{\mu}  \right)\nonumber \\
+ \int d^4x & \left( \frac{1}{2} S^{[ab]}_{\mu} (-\partial^2) S^{[ab]}_{\mu} +  \frac{1}{2} U^{[ab]}_{\mu}(-\partial^2)U^{[ab]}_{\mu} +  {\bar \vp}^{(ab)}_{\mu}(-\partial^2)\vp^{(ab)}_{\mu} \right) \;. \label{quad4}
\end{align}
The fields $\lambda^{a}_{\mu}, \eta^{a}_{\mu}$ describe the $i$-particles of the Gribov-Zwanziger action.

\subsection{Evaluating correlation functions in the Gribov-Zwanziger theory using $i$-particles}
Let us give here two examples of one loop correlation functions of composite operators
constructed from the $i$-fields. To this order, one can introduce the $i$-field strengths defined by
\begin{align}
\lambda^{a}_{\mu\nu} & = \partial_{\mu}  \lambda^{a}_{\nu} -  \partial_{\nu}  \lambda^{a}_{\mu} \;, \nonumber \\
\eta^{a}_{\mu\nu} & = \partial_{\mu}  \eta^{a}_{\nu} -  \partial_{\nu}  \eta^{a}_{\mu} \;. \label{fs}
\end{align}
The propagators are\footnote{There are more propagators than the ones
shown, but these are not relevant for the calculation presented.}
\begin{align}
\langle \lambda^{a}_{\mu}(k) \lambda^{b}_{\nu}(-k) \rangle & =  \frac{\delta^{ab}}{k^2  + i \hat{\gamma}^2} \left(\delta_{\mu\nu} -\frac{k_{\mu}k_{\nu}}{k^2} \right) \;, \nonumber \\
\langle \eta^{a}_{\mu}(k) \eta^{b}_{\nu}(-k) \rangle & = \frac{\delta^{ab}}{k^2  - i \hat{\gamma}^2} \left(\delta_{\mu\nu} -\frac{k_{\mu}k_{\nu}}{k^2} \right) \;, \label{letp}
\end{align}
with $\hat{\gamma}^4=2N\,g^2\,\gamma^4$ as introduced in Section \ref{sec:intro}. \\\\As simplest examples we investigate the following composite operators at leading order:
\begin{align}
O^{(1)}_{\lambda\eta}(x) &=  \left( \lambda^{a}_{\mu\nu}(x) \eta^{a}_{\mu\nu}(x) \right) \;, \nonumber  \\
O^{(2)}_{\lambda\eta}(x) &=  \varepsilon_{\mu\nu\rho\sigma} \left( \lambda^{a}_{\mu\nu}(x) \eta^{a}_{\rho\sigma(x)} \right) \;.
\end{align}
The integrals to be calculated for the correlation functions are
\begin{align}
\langle O^{(1)}_{\lambda\eta}(k) O^{(1)}_{\lambda\eta}(-k) \rangle & = (N^2-1) \int \frac{d^dp}{(2\pi)^d} \frac{4p^2(p-k)^2+4(d-2)(p^2-p\,k)^2}{(p^2-i\gam^2)((p-k)^2+i\gam^2)}  \;,\\
\langle O^{(2)}_{\lambda\eta}(k) O^{(2)}_{\lambda\eta}(-k) \rangle  &= (N^2-1) \int \frac{d^dp}{(2\pi)^d} \frac{32(k^2 p^2-(k\,p)^2)}{(p^2-i\gam^2)((p-k)^2+i\gam^2)} \;.
\end{align}
To avoid to dwell on technical details, let us here only outline the results for $d=2$. We have collected the technical details, including those for $d=4$, in Appendix \ref{app3}.
We are interested in
\begin{eqnarray}
\braket{ O^{(1)}_{\lambda\eta}(k) O^{(1)}_{\lambda\eta}(-k)} = 4 (N^2-1) F(k^2)\label{O1}\;,
\end{eqnarray}
with
\begin{eqnarray}
F(k^2)= \int\frac{d^dp}{(2\pi)^d} \frac{p^2(p-k)^2+(d-2)(p^2-p\,k)^2}{(p^2+i\gam^2)((p-k)^2-i\gam^2)}\;.  \label{O2}
\end{eqnarray}
After the introduction of a Feynman parameter and some manipulations, it turns out that we may write, see \eqref{O17b},
\begin{eqnarray}\label{O17}
&& F(k^2)-k^2\left[\frac{\p F(k^2)}{\p k^2}\right]_{k^2=0}-F(0)\nonumber\\&=& \frac{1}{4\pi}\int_1^{+\infty}d\tau\frac{1}{2\sqrt{\tau^2-1}}\frac{1}{k^2+\tau}+\frac{k^2}{4\pi}\int_1^{+\infty}d\tau\frac{1}{2\sqrt{\tau^2-1}}\frac{1}{\tau^2}- \frac{1}{4\pi}\int_1^{+\infty}d\tau\frac{1}{2\sqrt{\tau^2-1}}\frac{1}{\tau}\;.
\end{eqnarray}
We thus find
\begin{equation}\label{O18}
\rho(\tau)=\frac{1}{8\pi}\frac{1}{\sqrt{\tau^2-1}}\;.
\end{equation}
We conclude that, upon restoring units, we \emph{formally} have
\begin{equation}\label{O19}
\braket{ O^{(1)}_{\lambda\eta}(k) O^{(1)}_{\lambda\eta}(-k)}  =\int_{2\gam^2}^{+\infty}\frac{2 (N^2-1) \gam^4}{\pi\sqrt{\tau^2-4\gam^4}}\frac{d\tau}{\tau+k^2}\;.
\end{equation}
We clearly notice that the spectral density $\rho(\tau)$ is positive for $\tau\geq 2\gam^2$. The result as written in \eqref{O19} is indeed only formally correct, since the l.h.s.~of \eqref{O19} is divergent, directly seen upon inspection of its definition \eqref{O2}. Nevertheless, the spectral representation \eqref{O19} appearing in the r.h.s.~defines a finite function. The apparent contradiction is easily resolved by realizing that one should in fact refer to \eqref{O17}, which gives the correctly subtracted result. \\\\Let us now turn to the analysis of
\begin{eqnarray}
\braket{ O^{(2)}_{\lambda\eta}(k) O^{(2)}_{\lambda\eta}(-k)} = 32 (N^2-1) G(k^2)\label{P1}\;,
\end{eqnarray}
with
\begin{eqnarray}
G(k^2)= \int\frac{d^dp}{(2\pi)^d} \frac{k^2 p^2 - (k p)^2}{(p^2+i\gam^2)((p-k)^2-i\gam^2)}\;.  \label{P2}
\end{eqnarray}
We again refer to Appendix \ref{app3}. After the smoke clears, we find in \eqref{laatst}
\begin{eqnarray}
&& G(k^2)-G^2\left[\frac{\p G(k^2)}{\p k^2}\right]_{k^2=0}-G(0)\nonumber\\&=& \frac{1}{8\pi}\int_1^{+\infty}d\tau \sqrt{\tau^2-1}\frac{1}{k^2+\tau}+\frac{k^2}{8\pi}\int_1^{+\infty}d\tau\sqrt{\tau^2-1}\frac{1}{\tau^2}- \frac{1}{8\pi}\int_1^{+\infty}d\tau \sqrt{\tau^2-1} \frac{1}{\tau}\;.
\end{eqnarray}
The spectral density can be read off,
\begin{equation}
\rho(\tau)=\frac{1}{8\pi}\sqrt{\tau^2-1} \;,
\end{equation}
whereby $\rho(\tau)\geq0$ for $\tau\geq 1$.  We reintroduce the units, and we conclude that
\begin{equation}\label{P19}
\braket{ O^{(2)}_{\lambda\eta}(k) O^{(2)}_{\lambda\eta}(-k)}  =\frac{4(N^2-1)}{\pi}  \int_{2\gam^2}^{+\infty}d\tau\frac{\sqrt{\tau^2-4\gam^4}}{\tau+k^2}\;,
\end{equation}
which is again a formal result due to the divergent nature of both l.h.s.~and r.h.s.~.\\\\Using a similar analysis, one can also derive the spectral densities if $d=4$. The spectral densities in all cases turn out to be positive and thus at least at leading order the operators $O^{(1)}_{\lambda\eta}$ and $O^{(2)}_{\lambda\eta}$ appear to be physical.\\\\We conclude by remarking that the physical  meaning of the $i$-fields  $\lambda^{a}_{\mu}, \eta^{a}_{\mu}$ is akin to that of the $i$-fields $(\lambda,\eta)$ of the toy model. In fact, from equations (\ref{quad3}), (\ref{fv}) one easily gets
\begin{align}
\lambda^{a}_{\mu\nu}  = \frac{1}{\sqrt{2}} \left( 1 -
\frac{ig\gamma^2 \sqrt{2N}}{(-\partial^2)}  \right) (\partial_{\mu} A^{a}_{\nu} - \partial_{\nu}A^{a}_{\nu})
\;, \label{ifs1}
\end{align}
and a similar expression for $\eta^{a}_{\mu\nu}$.  As expected, the $i$-field strength   $\lambda^{a}_{\mu\nu}$ contains an explicit dependence from the nonperturbative Gribov parameter $\gamma$.  This dependence signals that, in order to construct  a sensible operator displaying good analyticity properties in the Gribov-Zwanziger theory, the starting gauge invariant operator obtained in the Faddeev-Popov theory, {\it i.e.}~without implementing the restriction to the Gribov region $\Omega$,  has to be supplemented by the addition of terms which are $\gamma$-dependent. To some extent, one might figure out that a  would-be physical operator in the presence of the Gribov horizon has to be constructed by deforming the starting gauge invariant operator by appropriate terms which exhibit an explicit dependence from the Gribov parameter $\gamma$, and which allow for a cancelation mechanism of the unphysical cuts appearing in expressions like those of eqs.(\ref{z2}), (\ref{z3}). Although the $i$-variables  $(\lambda^{a}_{\mu\nu}, \eta^{a}_{\mu\nu})$, introduced here in the quadratic approximation, enable us to  obtain in  a relatively simple way examples of correlation functions  possessing a K\"all\'{e}n-Lehmann spectral representation at one loop order, it is certainly tempting to look at those variables in the full Gribov-Zwanziger action, trying to unravel a systematic mechanism to take advantage of the presence of the Gribov horizon to cancel the unwanted unphysical cuts.

\section{Conclusions} \label{concl}

In this work we have pursued the investigation of the analyticity properties of correlation functions evaluated with a confining propagator of the Gribov type. In particular, as illustrated in the toy model, we have been able to characterize examples of composite operators whose correlation functions display cuts only on  the negative real axis, while possessing a positive spectral function, a result which has been extended to higher loop correlation functions. The introduction  of $i$-particles, which seem rather natural objects when dealing with a Gribov type propagator, has proven to be very useful  in the construction of such composite operators. \\\\In the case of the Gribov-Zwanziger theory, so far, we have been able to provide examples of composite operators, made of $i$-particles, whose correlation functions  at one loop order exhibit real cuts only. \\\\The introduction of $i$-fields in the full Gribov-Zwanziger action within a local and renormalizable framework is certainly a point worth to be investigated. This could open the possibility to obtain examples of correlation functions displaying good analyticity properties at higher orders, a result which can be regarded as a highly nontrivial achievement.  The systematic characterization of how a given gauge invariant operator needs to be deformed by the Gribov horizon in order to  possess a K{\"a}ll{\'e}n-Lehmann representation is a big challenge, requiring many ingredients as, for example, the mastering of the renormalization procedure of gauge invariant operators within the Gribov-Zwanziger action, a necessary step for a consistent higher order calculation. Let us remind here that, due to the soft breaking of the BRST symmetry, the issue of the renormalization of gauge invariant composite operators requires a careful analysis, as recently  done in \cite{Dudal:2009zh}, where a renormalization group invariant, and thus a fortiori finite operator in the Gribov-Zwanziger theory, which contains the scalar glueball operator $F^2(x)=F^{a}_{\mu\nu}(x) F^{a}_{\mu\nu}(x)$,  was already identified.  In particular, we point out that the soft breaking of the BRST symmetry implies that the operator $F^2$ mixes not only with BRST exact composite operators, but also with BRST non-invariant local quantities, which are determined by the softly broken Slavnov-Taylor identities \cite{Dudal:2009zh}. Evidently, the situation generalizes to higher dimensional  local composite operators containing three or more field strengths $F^{a}_{\mu\nu}$  which are contracted in such a way to give rise to a color singlet as,  e.g.~, $f^{abc} F^a_{\mu\nu} F^{b}_{\nu\rho} F^c_{\rho\mu}$.  Besides the mixing with other gauge invariant operators, a nontrivial mixing matrix between BRST exact terms as well as BRST non-invariant terms is to be expected here too.  This mixing should be taken into proper account, as (1) it is indispensable to obtain finite results, (2) it can play a major role in identifying the analyticity properties of the \emph{complete} operator, which differs from the classically gauge invariant one. From this perspective, the lowest order results presented for the Gribov-Zwanziger correlation functions \eqref{GZres1}, \eqref{GZres4}, \eqref{GZres2} and \eqref{GZres3} are only a first step. It will be interesting to find out whether one could also construct a renormalization group invariant operator in terms of the $i$-particles, and what the connection would be with the one already discussed in \cite{Dudal:2009zh}. \\\\As this paper only contains the first effort in constructing meaningful correlation functions in the presence of a Gribov type propagator \eqref{z11}, we did not yet study the more general gluon propagator
    \begin{equation}
\left\langle A_{\mu }^{a}(k)A_{\nu }^{b}(-k)\right\rangle =\delta
^{ab}\left( \delta _{\mu \nu }-\frac{k_{\mu }k_{\nu }}{k^{2}}\right) \frac{%
k^{2}+M^2}{k^{4}+M^2k^2+{\hat \gamma }^{4}}\;, \qquad {\hat{\gamma}}^4 = 2 g^2 N \gamma^4 \;, \label{z11bis}
\end{equation}%
which was discussed in \cite{Dudal:2008sp,Dudal:2007cw,Sorella:2009vt}, giving rise to the so called refined Gribov-Zwanziger action.  It  arises when additional nonperturbative quantum effects are considered. One might wonder if it would still be possible to obtain correlation functions with only physical cuts along the negative real axis when referring to the gluon propagator \eqref{z11bis}. One can easily write down the toy model analogy of the refined Gribov-Zwanziger action. It turns out that the previous question can be answered positively, again by introducing the $i$-particles. We refer to \cite{Dudal:2010b}, as the explicit evaluation of the spectral densities calls for a set of mathematical tools rather different from the ones used in the current paper. \\\\
Let us end this work by mentioning that several aspects related to the Gribov-Zwanziger theory as well as to the use of the Gribov type propagator and of the corresponding $i$-particles interpretation are still open. This is the case, for example, for the construction of the Minkowskian version of the theory, which remains to be achieved. We emphasize  that, as it stands, the Gribov-Zwanziger action has to be understood as an Euclidean field theory whose origin can be traced back to the lattice formulation of gauge theory. \\\\It is worth recalling here that gauge theory is well-defined on a finite lattice.  It is an Euclidean theory for which the correlation functions  of gauge non-invariant fields such as the gluon and quark fields vanish \cite{Seiler:1982pw}. The Osterwalder-Schrader reflection positivity \cite{Osterwalder:1973dx,Osterwalder:1974tc} holds for gauge invariant operators, so that for these quantities there exists a positive metric quantum mechanical Hilbert space \cite{Seiler:1982pw,Osterwalder:1977pc}.  Moreover, there exists a positive transfer matrix $T$, with $0 < T \leq 1$, which in the continuum limit would imply a positive Hamiltonian \cite{Osterwalder:1977pc,Seiler:1982pw,Luscher:1976ms}. In \cite{Zwanziger:1991ac},  it has been argued that the critical limit of lattice gauge theory, in the minimal Landau gauge, is precisely the theory we are discussing.  If so, the Gribov-Zwanziger theory
would inherit from lattice gauge theory a positive metric Hilbert space for composite gauge invariant quantities and a positive Hamiltonian, although in four dimensions this is not trivial because of renormalization. \\\\We also point out that the complex singularities exhibited by the Gribov propagator and by the $i$-particles might  jeopardize  the usual implementation of the Wick rotation in the correlation functions, so that the construction of the Minkowskian version of the theory is still an open point. In particular, as is apparent from the presence of complex poles, the two-point correlation function of the elementary Euclidean gluon field, as described by the $i$-particles, cannot be interpreted as a propagator of a physical particle in Minkowskian space, but may be appropriate for a confined gluon. On physical grounds, we expect that the rotation to Minkowskian space should be possible only for a restricted class of composite operators, which should be related to the physical spectrum of QCD. To some extent, this is precisely what emerges form the present analysis, namely: the K\"all\'{e}n-Lehmann representation seems to exist only for a restricted class of operators. The lack of a Minkowskian description of the $i$-particles and the presence of complex poles also indicate that the usual assumptions (or axioms) of local quantum field theory do not apply, in particular the spectral condition about the location of branch cuts for a two-point function \cite{Bogolyubov:1975ps}.
\\\\The Euclidean nature of the $i$-particles allows us to argue that they describe a phase for which the gluons are unphysical, \textit{i.e.}~ they do not appear as asymptotic states in the $S$-matrix elements. As such, the issue of the cancelation mechanism of the singularities in the $S$-matrix gluon amplitudes \cite{Habel:1989aq,Habel:1990tw,Stingl:1994nk,Driesen:1997wz,Driesen:1998xc}, which would arise due to the use of a propagator as \eqref{z11} or \eqref{z11bis}, is not addressed here. Evidently, we are still far from a concrete description, within a purely quantum field theory framework, of the mechanism which would account for the conversion of the gluon cloud into physical jets of hadronic matter\footnote{See also the related discussion about the possible interpretation of the gluon dispersion relation implied by the use of a  Gribov type propagator presented in \cite{Stingl:1985hx}.}.
\\\\In conclusion, although the results which we have obtained so far look promising, a satisfactory characterization of the analyticity properties of the correlation functions of the physical composite operators which would correspond to the spectrum of QCD remains to be achieved.

\section*{Acknowledgments}
D.~Dudal and N.~Vandersickel are supported by the Research-Foundation
Flanders (FWO Vlaanderen). S.~P.~Sorella is supported by the FAPERJ, Funda{%
\c c}{\~a}o de Amparo {\`a} Pesquisa do Estado do Rio de Janeiro, under the
program \textit{Cientista do Nosso Estado}, E-26/100.615/2007. The Conselho
Nacional de Desenvolvimento Cient\'{\i}fico e Tecnol\'{o}gico (CNPq-Brazil),
the Faperj, Funda{\c{c}}{\~{a}}o de Amparo {\`{a}} Pesquisa do Estado do Rio
de Janeiro, the SR2-UERJ and the Coordena{\c{c}}{\~{a}}o de Aperfei{\c{c}}%
oamento de Pessoal de N{\'{\i}}vel Superior (CAPES) are gratefully
acknowledged for financial support. M.~Q.~Huber is supported by the Doktoratskolleg ``Hadrons in Vacuum, Nuclei and Stars'' of the FWF under contract W1203-N08.
Furthermore he is indebted to the Instituto de F\'isica  of the Universidade do Estado do Rio de Janeiro, where part of this work was carried out.

\appendix
\section{K\"all\'{e}n-Lehmann representation of the toy model two-point function $\langle O_1(k) O_1(-k) \rangle$ for general $d$}\label{app1}
\subsection{Using the Schwinger parametrization}
In this Appendix, we shall verify that the two-point function $\langle O_1(k) O_1(-k) \rangle$, defined in eq.~\eqref{o1c}, has a spectral representation with associated positive spectral density for any value of $d=2,3,4$. We have already discussed the case $d=4$ in section \ref{subsubsec:D4}, but we shall follow a follow a different route here, which is applicable for general $d$, thereby serving as a check on the $d=4$ result we derived in expression \eqref{int12b}. Returning to expression (\ref{o1c}), we observe that by using the Schwinger parametrization, given in eqs.~(\ref{eq:Schwinger1}) and (\ref{eq:Schwinger2}), it can be cast  into the form
\begin{align}
\langle O_1(k) O_1(-k) \rangle &= \int \frac{d^dp}{(2\pi)^d}\; \frac{p^2(k-p)^2 + 2\theta^4}{((k-p)^4+2\theta^4)(p^4+2\theta^4)} \nonumber\\
&= \int^{\infty}_{0}d\alpha \int^{\infty}_{0}d\beta\; \cos\left( \sqrt{2}\theta^2(\alpha - \beta)\right) \int \frac{d^dp}{(2\pi)^d}\; e^{-\alpha p^2 - \beta(p-k)^2}\nonumber\\
&= \int^{\infty}_{0}d\alpha \int^{\infty}_{0}d\beta\; \cos\left( \sqrt{2}\theta^2(\alpha - \beta)\right)\frac{e^{-\frac{\alpha\beta}{\alpha+\beta}k^2}}{\left(4\pi(\alpha +\beta)\right)^{\frac{d}{2}}} \nonumber\\
&= \int^{\infty}_{0}d\alpha \int^{\infty}_{0}d\beta\; \mathrm{Re}\left\{ \frac{e^{-\left(\frac{\alpha\beta}{\alpha+\beta}k^2 - i \sqrt{2}\theta^2(\alpha - \beta)\right) } }{\left(4\pi(\alpha +\beta)\right)^{\frac{d}{2}}}\right\}   \;. \label{swg-par}
\end{align}
The cut structure is encoded in the exponential of the last line. In order to make this structure more apparent,  a few algebraic  manipulations of this expression are needed. Following  \cite{Zwanziger:1989mf},  as a first step we insert in (\ref{swg-par}) the unity decomposition
\begin{align}
1 = \int^{\infty}_{0}d\lambda \delta(\alpha + \beta - \lambda)  \;. \label{swg-unity}
\end{align}
After a rescaling $\alpha \rightarrow \lambda\alpha$, $\beta \rightarrow \lambda\beta$, the integral over $\lambda$ can be evaluated, yielding
\begin{align}
\langle O_1(k) O_1(-k) \rangle
&= \int^{1}_{0}d\alpha \int^{1}_{0}d\beta\delta(\alpha + \beta - 1)\int^{\infty}_{0}d\lambda \; \lambda^{1-\frac{d}{2}} \mathrm{Re}\left\{ \frac{e^{-\lambda\left(\frac{\alpha\beta}{\alpha+\beta}k^2 - i \sqrt{2}\theta^2(\alpha - \beta)\right) }}{\left(4\pi(\alpha +\beta)\right)^{\frac{d}{2}}}\right\} \nonumber\\
&= \frac{\Gamma\left(2-\frac{d}{2}\right)}{\left(4\pi\right)^{\frac{d}{2}}}\int^{1}_{0}d\alpha \int^{1}_{0}d\beta\delta(\alpha + \beta - 1)\; \mathrm{Re}\left\{  \left(\alpha\beta k^2 - i \sqrt{2}\theta^2(\alpha - \beta)\right)^{\frac{d}{2}-2}\right\}   \;. \label{swg-par2}
\end{align}
The second step is to parameterize the branch cut. For that purpose, exactly as before, we introduce the variable $u$:
\begin{align}
u\equiv \frac{\alpha -\beta}{2\alpha\beta} = \frac{2\alpha - 1}{2\alpha(1-\alpha)}  \;, \label{swg-s}
\end{align}
where the last equality follows from the constraint imposed by the delta function: $\beta = 1 - \alpha$. In terms of $u$, expression (\ref{swg-par2}) takes the form
\begin{align}
\langle O_1(k) O_1(-k) \rangle
= \frac{\Gamma\left(2-\frac{d}{2}\right)}{\left(4\pi\right)^{\frac{d}{2}}}\mathrm{Re}\left\{\int^{\infty}_{-\infty}du \frac{1}{\sqrt{1+u^2}}\left( \frac 12 \frac{1}{1 + \sqrt{1+u^2}}\right)^{\frac{d}{2} -1}\;   \left(k^2 - i 2\sqrt{2}\theta^2u\right)^{\frac{d}{2}-2}\right\}   \;. \label{swg-par-s}
\end{align}
Again we note that due to the first square-root in the integrand, this expression has a branch cut in the complex $u$-plane which starts at $u=i$ and extends till $u=i\infty$. This cut may be rotated to the real axis by exploiting the properties of the contour integral to express the result as an integral along the cut exactly as we did before.  Defining $u=iy$ we get
\begin{align}
\langle O_1(k) O_1(-k) \rangle
= \frac{4\Gamma\left(2-\frac{d}{2}\right)}{\left(8\pi\right)^{\frac{d}{2}}}\mathrm{Re}\left\{\int^{\infty}_{1}dy \frac{1}{\sqrt{y^2-1}} \left( \frac{1}{1 +
i \sqrt{y^2-1}}\right)^{\frac{d}{2} -1}\;   \left(k^2 + 2\sqrt{2}\theta^2y\right)^{\frac{d}{2}-2}\right\}   \;. \label{swg-par-y}
\end{align}
For the case $d=2$, we can infer the spectral density directly from \eqref{swg-par-y}. We find
\begin{align}
\langle O_1(k) O_1(-k) \rangle
= \frac{1}{2\pi}\int_{2\sqrt{2}\theta^2}^{+\infty}\frac{d\tau}{\sqrt{\tau^2-8\theta^4}}\frac{1}{k^2+\tau} \;, \label{new1}
\end{align}
hence
\begin{align}
\rho_{d=2}(\tau) = \frac{1}{2\pi\sqrt{\tau^2-8\theta^4}} \;, \label{new2}
\end{align}
with indeed $\rho_{d=2}(\tau)\geq0$ for $\tau\geq 2\sqrt{2}\theta^2$. Explicitly, the two-point function yields
\begin{align}
\langle O_1(k) O_1(-k) \rangle=\frac{\arccos\left(\frac{k^{2}}{2\sqrt{2}\theta^2}\right)}{2\pi\sqrt{8\theta^4-k^{4}}} \;, \label{new3}
\end{align}
displaying a branch cut from $-\infty$ to $-2\sqrt{2}\theta^2$.\\\\
For a general $d$, it is useful to employ the following identity
\begin{align}
\left(k^2 + 2\sqrt{2}\theta^2y\right)^{\frac{d}{2}-2} = \frac{1}{\Gamma\left(2-\frac{d}{2}\right)\Gamma\left(\frac{d}{2}-1\right)}\int^{\infty}_{2\sqrt{2}\theta^2y}d\tau\frac{\left(\tau - 2\sqrt{2}\theta^2y\right)^{\frac{d}{2}-2}}{\tau+k^2}   \;, \label{swg-ident}
\end{align}
which already displays the familiar structure found in the K\"all\'{e}n-Lehmann representation. Inserting this identity in (\ref{swg-par-y}), it is not difficult to see that the resulting overall range of integration can be reordered as
\begin{align}
\int^{\infty}_{1}dy \int^{\infty}_{2\sqrt{2}\theta^2y}d\tau \left(\cdots\right) = \int^{\infty}_{2\sqrt{2}\theta^2}d\tau \int^{\frac{\tau}{2\sqrt{2}\theta^2}}_{1}dy \left(\cdots\right)\;. \label{swg-range}
\end{align}
We make now a further change of variables defining
\begin{align}
\frac 1y \equiv \cos\gamma; \;\; 0\leq\gamma\leq\frac \pi2 \;. \label{swg-change-var}
\end{align}
Putting all together, it is just a matter of algebraic manipulation to obtain
\begin{align}
\langle O_1(k) O_1(-k) \rangle
= \int^{\infty}_{2\sqrt{2}\theta^2}d\tau \frac{\rho(\tau)}{\tau+k^2} \;, \label{swg-KL}
\end{align}
where the spectral function $\rho(\tau)$ is given by
\begin{align}
\rho(\tau) = \frac{4\tau^{\frac{d}{2}-2}}{\left(8\pi\right)^{\frac{d}{2}}\Gamma\left(\frac{d}{2}-1\right)}\int^{\phi(\tau)}_{0}d\gamma\; \cos\left(\left(\frac{d}{2}-1\right)\gamma\right)\;   \left(\cos\gamma -\cos\phi(\tau)\right)^{\frac{d}{2}-2}  \;, \label{swg-espec}
\end{align}
with $\phi(\tau)$ defined as
\begin{align}
\cos\phi(\tau) \equiv \frac{2\sqrt{2}\theta^2}{\tau}\;. \label{swg-phi}
\end{align}
For $d=4$, the spectral function can be immediately evaluated
\begin{align}
\rho_{d=4}(\tau) = \frac{1}{\left(4\pi\right)^{2}}\sin\phi(\tau) = \frac{1}{\left(4\pi\right)^{2}} \sqrt{1-\frac{8\theta^4}{\tau^2}}   \;, \label{swg-especD4}
\end{align}
which is consistent with \eqref{int12b}.\\\\When $d=3$, we need to evaluate
\begin{equation}\label{int15}
    \rho_{d=3}(\tau)=\frac{1}{4\sqrt{2}\pi^2}\frac{1}{\sqrt{\tau}}\int_0^\phi d\gamma \frac{\cos(\gamma/2)}{\sqrt{\cos\gamma-\cos\phi}}\;.
\end{equation}
We must compute an integral of the class
\begin{eqnarray}\label{int16}
    \int_0^\phi\frac{\cos^{2n+1}(\gamma/2)d\gamma}{\sqrt{\cos\gamma-\cos\phi}}&=& 2\int_0^{\phi/2}\frac{\cos^{2n+1}(\gamma')d\gamma'}{\sqrt{1-2\sin^2\gamma'-\cos\phi}}=2\int_0^{\sin(\phi/2)}\frac{(1-\beta^2)^nd\beta}{\sqrt{1-\cos\phi-2\beta^2}}\,,
\end{eqnarray}
with $n\in\mathbb{N}$, where we used the substitution $\sin \gamma'=\beta$. This integral \eqref{int16} can be expressed in terms of the Gaussian hypergeometric function,
\begin{eqnarray}
   2\int_0^{\sin(\phi/2)}\frac{(1-\beta^2)^nd\beta}{\sqrt{1-\cos\phi-2\beta^2}}=\frac{\pi}{\sqrt{2}}\,{}_2 F_1(1/2;-n;1,\sin^2(\phi/2))\,.
\end{eqnarray}
Since, in particular ${}_2 F_1(1/2;0;1,x)= 1$, we obtain
\begin{equation}\label{int18-1}
    \rho_{d=3}(\tau)=\frac{1}{8\pi}\frac{1}{\sqrt{\tau}}\,,
\end{equation}
also positive for $\tau\geq 2\sqrt{2}\theta^2$. As such,
\begin{equation}\label{int18-2}
    F(k^2)=\int_{2\sqrt{2}\theta^2}^{\infty}d\tau\frac{\rho(\tau)}{\tau+k^2}=\frac{1}{4\pi}\frac{1}{k^2}\mathrm{arctan}\sqrt{\frac{2\sqrt{2}\theta^2}{k^2}}\,,
\end{equation}
for $d=3$, valid in the complex plane, exhibiting a branch cut for $k^2\in[-\infty,-2\sqrt{2}\theta^2]$.

\subsection{Using hypergeometric functions}
A very convenient method to evaluate Feynman diagrams is the use of hypergeometric functions, since these are meromorphic functions in the complex plane. The matter of analytic continuation is especially clear, when one uses these functions as will become evident in the following calculation. Another advantage is the possibility of keeping more parameters general like the number of dimensions or the exponent of the propagator, which can even be non-integer as needed sometimes. Hypergeometric functions have proven useful especially for higher n-point functions or diagrams with many mass scales, because they allow to write down the result as a closed expression.\\\\
One possibility to arrive at hypergeometric functions is the use of the negative dimension integration method (NDIM) \cite{Dunne:1987am,Halliday:1987an,Dunne:1987qb,Ricotta:1990nd}, but one can for example also work with Mellin-Barnes representations \cite{Berends:1993ee} or dispersion relations \cite{Bauberger:1994by}. The result for the one loop two-point diagram with masses\footnote{The choice of the square root of two for the masses is in correspondence to the main part of the  paper.} $m^{2}:=m_{1}^{2}=i\,\sqrt{2}\theta^{2}$ and $m_{2}^{2}=-i\,\sqrt{2}\theta^{2}$ in $d$ dimensions with exponents $i_1$ and $i_2$ for the propagators is \cite{Anastasiou:1999ui}
\begin{align}
I_{2}(i_{1},i_{2};m_{1},m_{2};d)&=\int\frac{d^{d}q}{(2\pi)^{d}}((k+q)^{2}+m_{1}^{2})^{i_{1}}(q^{2}+m_{2}^{2})^{i_{2}}\nnnl
&=  (4\pi)^{-d/2}(-1)^{i_{1}+i_{2}}\Bigg((m_{1}^{2})^{d/2+i_{1}+i_{2}}\frac{(d/2,i_{2})}{(-d/2-i_{2}-i_{2},d/2+i_{2})}\nnnl
 & \times F_{4}(-i_{1}-i_{2}-d/2,-i_{2};1-i_{2}-d/2,d/2;\frac{m_{2}^{2}}{m_{1}^{2}},\frac{k^{2}}{m_{1}^{2}})\nnnl
+ & (m_{1}^{2})^{i_{1}}(m_{2}^{2})^{d/2+i_{2}}(-i_{2},-d/2)F_{4}(-i_{2},d/2;1+d/2+i_{2},d/2;\frac{m_{2}^{2}}{m_{2}^{2}},\frac{-k^{2}}{m_{1}^{2}})\Bigg),
\end{align}
where the Pochhammer symbol is defined as
\begin{align}
 (a,n)=\frac{\Gamma(a+n)}{\Gamma(a)}.
\end{align}
$F_4$ is the fourth Appell function, which is a meromorphic function of two variables.
At this point we will choose a specific series representation for it:
\begin{align}\label{eq:F4-series}
F_{4}(a,b;c,d;x,y)=\sum_{m,n=0}^{\infty}\frac{(a,m+n)(b,m+n)}{(c,m)(d,n)}\frac{x^{m}}{m!}\frac{y^{n}}{n!}.
\end{align}
This restricts the validity of the following manipulations to a certain region of convergence, namely $\sqrt{|x|}+\sqrt{|y|}<1$. The fact that our masses are complex fits quite naturally in this framework, as the variables can be complex numbers. Later on we will switch back to hypergeometric \textit{functions}, where we do not have to worry about regions of convergence, since we can always choose a series representation appropriate for the values of the variables $x$ and $y$.\\\\
We continue by rewriting the Appell function into a Gaussian hypergeometric series,
\begin{align}
 F_{4}(a,b;c,d;x,y)=\sum_{n=0}^{\infty}\frac{(a,n)(b,n)}{(c,n)}\frac{x^{n}}{n!}\,_{2}F_{1}(a+n,b+n;d;y),
\end{align}
and evaluating the one-dimensional series with
\begin{align}
 _{2}F_{1}(a,b;a+1-b;-1)=2^{-a}\pi^{1/2}\frac{\Gamma(a+1-b)}{\Gamma(1/2+a/2)\Gamma(1+a/2-b)}.
\end{align}
We arrive at
\begin{align}
 I_{2}&(-1,-1;m_{1},e^{i\,\pi/2}m_{1};d)\nnnl
=&(4\pi)^{-d/2}\pi^{1/2}\Bigg((m_{1}^{2})^{d/2-2}2^{-2+d/2}\frac{(d/2,-1)\Gamma(2-d/2)}{(-d/2+2,d/2-1)\Gamma(3/2-d/4)\Gamma(1-d/4)}\nnnl
&\sum_{n=0}^{\infty}\frac{(-d/2+2,n)(1,n)}{(d/2,n)}\frac{1}{(3/2-d/4,n/2)(1-d/4,-n/2)}\left(\frac{-k^{2}}{2m^{2}}\right)^{n}\frac{1}{n!}\nnnl&+(m_{1}^{2})^{-1}(m_{2}^{2})^{d/2-1}2^{-d/2}\frac{(1,-d/2)\Gamma(d/2)}{\Gamma(1/2+d/4)\Gamma(d/4)}\nnnl
&\times\sum_{n=0}^{\infty}\frac{(1,n)(d/2,n)}{(d/2,n)}\frac{1}{(1/2+d/4,n/2)(d/4,-n/2)}\left(\frac{-k^{2}}{2m^{2}}\right)^{n}\frac{1}{n!}\Bigg).
\end{align}
To deal with the half-integer summation indices, we split the sum into even and odd indices:
\begin{align}\label{eq:2pt-final-d}
 I_{2}&(-1,-1;m_{1},e^{i\,\pi/2}m_{1};d)\nnnl
=&(4\pi)^{-d/2}\pi^{1/2}\Bigg((m_{1}^{2})^{d/2-2}2^{-2+d/2}\frac{(d/2,-1)\Gamma(2-d/2)}{(-d/2+2,d/2-1)\Gamma(3/2-d/4)\Gamma(1-d/4)}\nnnl
&\times\Big(\sum_{n=0}^{\infty}\frac{(-d/2+2,2n)(1,2n)}{(d/2,2n)}\frac{1}{(3/2-d/4,n)(1-d/4,-n)}\left(\frac{-k^{2}}{2m^{2}}\right)^{2n}\frac{1}{(2n)!}\nnnl
&+\sum_{n=0}^{\infty}\frac{(-d/2+2,2n+1)(1,2n+1)}{(d/2,2n+1)}\frac{1}{(3/2-d/4,n+1/2)(1-d/4,-n-1/2)}\left(\frac{-k^{2}}{2m^{2}}\right)^{2n+1}\frac{1}{(2n+1)!}\big)\nnnl
&+(m_{1}^{2})^{-1}(m_{2}^{2})^{d/2-1}2^{-d/2}\frac{1}{4}\frac{(1,-d/2)\Gamma(d/2)}{\Gamma(1/2+d/4)\Gamma(d/4)}\nnnl
&\times\Big(\sum_{n=0}^{\infty}\frac{(1,2n)(d/2,2n)}{(d/2,2n)}\frac{1}{(1/2+d/4,n)(d/4,-n)}\left(\frac{-k^{2}}{2m^{2}}\right)^{2n}\frac{1}{(2n)!}\nnnl
&+\sum_{n=0}^{\infty}\frac{(1,2n+1)(d/2,2n+1)}{(d/2,2n+1)}\frac{1}{(1/2+d/4,n+1/2)(d/4,-n-1/2)}\left(\frac{-k^{2}}{2m^{2}}\right)^{2n+1}\frac{1}{(2n+1)!}\big)\Bigg)\nnnl
=&(4\pi)^{-d/2}\pi^{1/2}\Bigg((m_{1}^{2})^{d/2-2}2^{-2+d/2}\frac{(d/2,-1)\Gamma(2-d/2)}{(-d/2+2,d/2-1)\Gamma(3/2-d/4)\Gamma(1-d/4)}\nnnl
&\times\Big(\,_{2}F_{1}(1,1-d/4;1/2+d/4;-\frac{k^{4}}{4m^{4}})\nnnl
&-\frac{k^{2}}{2m^{2}}\frac{(-d/2+2,1)}{(d/2,1)(3/2-d/4,1/2)(1-d/4,-1/2)}\,_{2}F_{1}(1,3/2-d/4;1+d/4;-\frac{k^{4}}{4m^{4}})\big)\nnnl
&+(m_{1}^{2})^{-1}(m_{2}^{2})^{d/2-1}2^{-d/2}\frac{(1,-d/2)\Gamma(d/2)}{\Gamma(1/2+d/4)\Gamma(d/4)}\Big(\,_{2}F_{1}(1,1-d/4;1/2+d/4;-\frac{k^{4}}{4m^{4}})\nnnl
&-\frac{k^{2}}{2m^{2}}\frac{1}{(1/2+d/4,1/2)(d/4,-1/2)}\,_{2}F_{1}(1,3/2-d/4;1+d/4;-\frac{k^{4}}{4m^{4}})\Big)\Bigg),
\end{align}
where the following identities for Pochhammer symbols  were used:
\begin{align}
 (a,m+n)=&(a,m)(a+m,n),\\
(a,2n)=&2^{2n}(a/2,n)(1/2+a/2,n),\\
\Gamma(a+m)=&\Gamma(a)(a,m),\\
(a,-n)=&\frac{(-1)^{-n}}{(1-a,n)},\quad n\,\text{integer}.
\end{align}
For specific values of $d$ the expression will simplify.\\\\Here we are dealing again with a hypergeometric function. We know that $_2F_1(a,b;c;x)$ is analytic in the complex plane except at $+1$, from where a branch cut to $+\infty$ starts\footnote{Strictly speaking this is only true for nonexceptional parameters $a$, $b$ and $c$; these are parameters for which the series terminates or the series is undefined.}. From the argument of $_2F_1$, it looks as if the result has two branch cuts: From $\pm2 \sqrt{2}\theta^2$ to $\pm \infty$. We will see below in the expressions for specific dimensions that only one of those remains, namely the physical one on the negative real axis.\\\\Now we want to obtain specific expressions in two, three and four dimensions. In two dimensions we get
\begin{align}
 I_{2}&(-1,-1;m_{1},e^{i\,\pi/2}m_{1};2-2\varepsilon)\nonumber\\=&(4\pi)^{-1+\varepsilon}\pi^{1/2}\Bigg((m_{1}^{2})^{-1+\varepsilon}2^{-1-\varepsilon}\frac{(1-\varepsilon,-1)\Gamma(1+\varepsilon)}{(1+\varepsilon,1-\varepsilon)\Gamma(1+\varepsilon/2)\Gamma(1/2+\varepsilon/2)}\times\nonumber\\&\times\Big(\,_{2}F_{1}(1,1/2+\varepsilon/2;1-\varepsilon/2;-\frac{k^{4}}{4m^{4}})+\nonumber\\&+\frac{k^{2}}{2m^{2}}\frac{(1+\varepsilon,1)}{(1-\varepsilon,1)(1+\varepsilon/2,1/2)(\varepsilon/2,-1/2)}\,_{2}F_{1}(1,1+\varepsilon/2;3/2-\varepsilon/2;-\frac{k^{4}}{4m^{4}})\big)+\nonumber\\&+(m_{1}^{2})^{-1}(m_{2}^{2})^{d/2-1}2^{-d/2}\frac{(1,-1+\varepsilon)\Gamma(-1+\varepsilon)}{\Gamma(1-\varepsilon/2)\Gamma(1/2-\varepsilon/2)}\Big(\,_{2}F_{1}(1,1/2+\varepsilon/2;1-\varepsilon/2;-\frac{k^{4}}{4m^{4}})+\nonumber\\&+\frac{k^{2}}{2m^{2}}\frac{1}{(1-\varepsilon/2,1/2)(1/2-\varepsilon/2,-1/2)}\,_{2}F_{1}(1,1+\varepsilon/2;3/2-\varepsilon/2;-\frac{k^{4}}{4m^{4}})\Big)\Bigg).
\end{align}
Although it is not evident from above due to divergent coefficients, the integral is finite in two dimensions. Performing the limit $\epsilon\rightarrow 0$ yields
\begin{align}
 I_{2}&(-1,-1;m_{1},e^{i\,\pi/2}m_{1};2)=\frac{\pi/2-  \arcsin\left(\frac{k^{2}}{2\sqrt{2}\theta^2}\right)}{2\pi\sqrt{8\theta^4-k^{4}}}=\frac{\arccos\left(\frac{k^{2}}{2\sqrt{2}\theta^2}\right)}{2\pi\sqrt{8\theta^4-k^{4}}}.
\end{align}
The $\arccos$ has branch cuts from $\pm1$ to $\pm \infty$ and the inverse square root from $0$ to $-\infty$. They combine in such a way that only a cut on the negative real axis starting at $-2\sqrt{2}\theta^2$ remains.\\\\For three dimensions the limit $\eps \rightarrow 0$ yields
\begin{align}
 I_{2}&(-1,-1;m_{1},e^{i\,\pi/2}m_{1};3)=\frac{6\sqrt{2}\theta^{2}\:_{2}F_{1}(\frac{1}{4},1,\frac{5}{4},\frac{k^{4}}{8\theta^{4}})-k^{2}\:_{2}F_{1}(\frac{3}{4},1,\frac{7}{4},\frac{k^{4}}{8\theta^{4}})}{48\pi\theta^{3}}.
\end{align}
With the integral representation for the Gaussian hypergeometric series,
\begin{align}\label{eq:int-rep}
 _{2}F_{1}(a,b,c,x)=&\frac{\Gamma(c)}{\Gamma(b)\Gamma(c-b)}\int_0^1 dt\, t^{b-1}(1-t)^{c-b-1}(1-t\, x)^{-a},\quad \mathrm{Re}\, c>\mathrm{Re}\, b>0,
\end{align}
this can be rewritten into a spectral representation:
\begin{align}\label{m1}
 I_{2}&(-1,-1;m_{1},e^{i\,\pi/2}m_{1};3)=\frac{1}{8\pi}\int_{2\sqrt{2}\theta^{2}}^{\infty}d\tau\frac{\tau^{-1/2}}{\tau+k^{2}}=\frac{\arctan(\sqrt{\frac{k^{2}}{2\sqrt{2}\theta^{2}}})}{4\pi\sqrt{k^{2}}}.
\end{align}
As the inverse tangent has branch cuts from $\pm i$ to $\pm i \infty$, we again observe a branch cut on the negative real axis. The role of the square root in the denominator is to compensate the cut stemming from the square root inside the  $\textrm{arctan}$, which would start at $k^2=0$. We notice the expression \eqref{m1} coincides with \eqref{int18-2}.\\\\Finally we evaluate \eref{eq:2pt-final-d} in four dimensions. Here more care is necessary since the integral is divergent:
\begin{align}\label{eq:I2-d4}
 I_{2}&(-1,-1;m_{1},e^{i\,\pi/2}m_{1};4-2\epsilon)\nnnl
=&(4\pi)^{-2+\epsilon}\pi^{1/2}\Bigg(\Big((m_{1}^{2})^{-\epsilon}2^{-\epsilon}\frac{\Gamma(1-\epsilon)\Gamma(\epsilon)\Gamma(\epsilon)}{\Gamma(2-\epsilon)\Gamma(1/2+\epsilon/2)\Gamma(\epsilon/2)}-(m_{2}^{2})^{-\epsilon}2^{\epsilon}\frac{1}{4}\frac{\Gamma(\epsilon-1)\Gamma(2-\epsilon)}{\Gamma(3/2-\epsilon/2)\Gamma(1-\epsilon/2)}\Big)\nnnl
&\times\,_{2}F_{1}(\epsilon/2,1;3/2-\epsilon/2;-\frac{k^{4}}{4m^{4}})\nnnl
&-\frac{k^{2}}{2m^{2}}(m_{2}^{2})^{-\epsilon}2^{\epsilon}\frac{1}{4}\Big(\frac{\Gamma(\epsilon-1)\Gamma(2-\epsilon)}{\Gamma(3/2-\epsilon/2)\Gamma(1-\epsilon/2)}\frac{\Gamma(\epsilon+1)\Gamma(2-\epsilon)\Gamma(1/2+\epsilon/2)\Gamma(\epsilon/2)}{\Gamma(\epsilon)\Gamma(3-\epsilon)\Gamma(1+\epsilon/2)\Gamma(-1/2+\epsilon/2)}\nnnl
&-\frac{\Gamma(\epsilon-1)\Gamma(2-\epsilon)}{\Gamma(3/2-\epsilon/2)\Gamma(1-\epsilon/2)}\frac{\Gamma(1-\epsilon/2)\Gamma(3/2-\epsilon/2)}{\Gamma(2-\epsilon/2)\Gamma(1/2-\epsilon/2)}\Big)\nnnl
&\times\,_{2}F_{1}(1,1/2+\epsilon/2;2-\epsilon/2;-\frac{k^{4}}{4m^{4}})\Bigg).
\end{align}
The coefficient of the second hypergeometric function and the function itself are finite for $\eps \rightarrow 0$ and only the first part yields divergences. To extract one factor of $\eps$ we can use the integral representation of the Gaussian hypergeometric function, \eref{eq:int-rep}. For the given values of the parameters ($a=1$, $b=\eps/2$, $c=3/2-\eps/2$) there is a pole at $t=0$ in the integrand. Rewriting the integral as
\begin{align}\label{eq:2F1-int-1}
 _{2}F_{1}(1,b,c,x)=&\frac{\Gamma(c)}{\Gamma(b)\Gamma(c-b)}\int_0^1 dt\, t^{b-1}(1-t)^{c-b-1}(1-t\, x)^{-1}=
\frac{\Gamma(c)}{\Gamma(b)\Gamma(c-b)}\int_0^1 dt\, t^{b-1}(1-t)^{c-b-1}f(t)=J_1+J_2\,,
\end{align}
with
\begin{align}
J_1&=\frac{\Gamma(c)}{\Gamma(b)\Gamma(c-b)}\int_0^1 dt\, t^{b-1}(1-t)^{c-b-1}f(0),\\
J_2&=\frac{\Gamma(c)}{\Gamma(b)\Gamma(c-b)}\int_0^1 dt\, t^{b-1}(1-t)^{c-b-1}(f(t)-f(0))\,,
\end{align}
it can be exposed explicitly. For the last term we calculate
\begin{align}
 f(t)-f(0)=t\frac{x}{1-t\,x}=t\,g(t).
\end{align}
This raises the exponent of  $t$ in $J_2$ by one. The integral in $J_1$ is identified as the $\beta$-function,
\begin{align}
 B(x,y)=\int_0^1 dt \,t^{x-1} (1-t)^{y-1}=\frac{\Gamma(x) \Gamma(y)}{\Gamma(x+y)}, \quad \mathrm{Re} \,x>1,\,\, \mathrm{Re}\,y>1.
\end{align}
Thus it yields $J_1=1$. Note that the conditions for this identity coincide with those of the integral representation of $_2F_1$. The second integral is written back into a Gaussian hypergeometric function:
\begin{align}\label{eq:1L-d4}
 J_2&=\frac{\Gamma(c)}{\Gamma(b)\Gamma(c-b)}\int_0^1 dt\, t^{b}(1-t)^{c-b-1}\frac{x}{1-t\,x}=x\frac{\Gamma(c)}{\Gamma(b)\Gamma(c-b)}\frac{\Gamma(b+1)\Gamma(c-b)}{\Gamma(c+1)}\,_2F_1(1,b+1;c+1;x)\nnnl
&=\frac{b}{c}\,x\,_2F_1(1,b+1;c-b;x)\quad \rightarrow \quad -\frac{k^{4}}{4m^{4}} \frac{\eps/2}{3/2-\eps/2} \,_2F_1(1,\eps/2+1;5/2-\eps/2;-\frac{k^{4}}{4m^{4}}).
\end{align}
The coefficient of this function in \eref{eq:I2-d4} has a term of order $1/\eps$. Thus we need the total expression in \eref{eq:1L-d4} up to order $\eps$. The goal in rewriting the original Gaussian hypergeometric function like this was to extract one factor of $\eps$, because now we need only the lowest order in $\eps$ of the Gaussian function. Combining all intermediary results we obtain
\begin{align}
J_{2}&(-1,-1;m_{1},e^{i\,\pi/2}m_{1};4-2\epsilon)= \frac{1}{16\pi^{2}}\frac{1}{\varepsilon}+\frac{2-\gamma+\ln(4\pi)-\ln \sqrt{2}\theta^2}{16\pi^{2}}+\frac{-\pi\sqrt{2} \theta^2+\sqrt{8\theta^4-k^{4}}\arccos(k^{2}/(2\sqrt{2}\theta^2))}{16\pi^{2}k^{2}}.
\end{align}
Again the cut the on the positive real axis is canceled by the one from the square root. The subtracted result is
\begin{align}
J_{2}&(-1,-1;m_{1},e^{i\,\pi/2}m_{1};4-2\epsilon)= \frac{k^2-\pi\sqrt{2} \theta^2+\sqrt{8\theta^4-k^{4}}\arccos(k^{2}/(2\sqrt{2}\theta^2))}{16\pi^{2}k^{2}}\,,
\end{align}
in accordance with \eqref{int12}.\\\\The explicit example of the one loop case demonstrates that hypergeometric functions are a useful method, which can develop its full power especially for higher loop cases. The derivation for an $N$-loop diagram in terms of generalized Lauricella functions can be found in \cite{Berends:1993ee}. For diagrams with only two massive propagators only two hypergeometric functions remain that can be treated along the lines above. The end result is again a cut only on the negative real axis.

\section{Spectral density for one real mass and one vanishing mass}\label{app2}
We start from
\begin{eqnarray}
F(k^2) &=& \int \frac{d^d p}{(2\pi)^d} \frac{1}{(k-p)^2 + m^2} \frac{1}{p^2}\,.
\end{eqnarray}
For $k^2>0$, we can employ the Feynman trick and differentiating w.r.t.~$k^2$ yields for $d=4$
\begin{eqnarray}
\frac{\p F(k^2)}{\p k^2} &=& - \frac{1}{16\pi^2}  \int_0^1 dx  \frac{x (1-x)}{x (1-x) k^2  + x m^2} \nonumber\\
&=& - \frac{1}{16\pi^2}  \int_0^1 dx  \frac{1}{ k^2  +  \frac{m^2}{1-x}}\,.
\end{eqnarray}
We perform a transformation of variables, by setting $s = \frac{m^2}{1-x}$,
\begin{eqnarray}
\frac{\p F(k^2)}{\p k^2} &=&  - \frac{m^2}{16\pi^2}  \int_{m^2}^{+\infty} ds  \frac{1}{s^2} \frac{1}{ k^2  +   s  } \nonumber\\
&=& + \frac{1}{16\pi^2}  \int_{m^2}^{+\infty} ds  \frac{d }{ ds} \left( \frac{m^2}{s}\right) \frac{1}{ k^2  +   s  }\,.
\end{eqnarray}
After doing a partial integration, we obtain
\begin{eqnarray}
\frac{\p F(k^2)}{\p k^2} &=&  \frac{1}{16\pi^2} \left[  \left. \frac{1}{ k^2  +   s  } \frac{m^2}{s} \right|_{s = m^2}^{+\infty} -  \int_{m^2}^{+\infty} ds \frac{m^2}{s}  \frac{d }{ ds} \left( \frac{1}{ k^2  +   s  }\right) \right] \nonumber\\
&=& \frac{-1}{16\pi^2} \left[  \frac{1}{ k^2  +   m^2 }   +  \int_{m^2}^{+\infty} ds \frac{m^2}{s}  \frac{-1}{( k^2  +   s )^2 }\right] \nonumber\\
&=&  \frac{-1}{16\pi^2}  \frac{\p }{\p k^2}   \left[  \ln ( k^2  +   m^2 )  +    \int_{m^2}^{+\infty} ds \frac{m^2}{s}\frac{1}{ k^2  +   s  }  \right]\,,
\end{eqnarray}
so that
\begin{eqnarray}
 F(k^2) - F(0) &=&    \frac{1}{16\pi^2}  \left[ - \ln ( k^2  +   m^2 ) + \ln m^2  -   \int_{m^2}^{+\infty} ds \frac{m^2}{s}\frac{1}{ k^2  +   s  }   + \int_{m^2}^{+\infty} ds \frac{m^2}{s}\frac{1}{ s  }  \right]  \nonumber\\
 &=&    \frac{1}{16\pi^2}  \left[  \int_{m^2}^{+\infty} ds \frac{1}{k^2 +s}-  \int_{m^2}^{+\infty} ds \frac{1}{s} -   \int_{m^2}^{+\infty} ds \frac{m^2}{s}\frac{1}{ k^2  +   s  }   + \int_{m^2}^{+\infty} ds \frac{m^2}{s}\frac{1}{ s  }  \right] \nonumber\\
 &=&    \frac{1}{16\pi^2}   \int_{m^2}^{+\infty} ds  \left[ \frac{1}{k^2 +s}- \frac{1}{s}\right]\left( 1-   \frac{m^2}{s}\right)\,.
\end{eqnarray}
We conclude that the spectral density is given by
\begin{eqnarray}\label{appspec}
\rho_1(s) &=& 1-   \frac{m^2}{s}\,,
\end{eqnarray}
which is indeed positive for $s\geq m^2$.

\section{Derivations of the spectral density for the simplest correlation functions in the Gribov-Zwanziger theory }\label{app3}
\subsection{Using the Feynman parametrization for $d=2$}
We wish to bring
\begin{eqnarray}
\braket{ O^{(1)}_{\lambda\eta}(k) O^{(1)}_{\lambda\eta}(-k)} = 4 (N^2-1) F(k^2)\label{O1b}\;,
\end{eqnarray}
into a spectral form, with $F(k^2)$ defined in eq.~\eqref{O2}. We first derive a Feynman parametrization of \eqref{O2}. Proceeding in the usual way one finds
\begin{eqnarray}
F(k^2)= \int_0^1 dx \int\frac{d^d q}{(2\pi)^d} \frac{N(q,k,x)}{(q^2+\Delta^2)^2}\;,  \label{O3}
\end{eqnarray}
where we used the substitution $q=p-kx$, and whereby
\begin{equation}
\Delta^2=x(1-x)k^2-(2x-1)i\gam^2\;.
\end{equation}
We shall temporarily work in units $2\gam^2=1$. We still have to identify the numerator $N(q,k,x)$. Keeping in mind that terms odd in $q_\mu$ will vanish upon integration, and that we may replace $q_\mu q_\nu\to q^2\frac{\delta_{\mu\nu}}{d}$ within the $q$-integral, we are brought to
\begin{eqnarray}
F(k^2)= (d-1)\int_0^1 dx \int\frac{d^d q}{(2\pi)^d} \frac{x^2(1-x)^2 k^4 + \frac{2}{d}\left[1-(d+2)x(1-x)\right]k^2q^2 + q^4}{(q^2+\Delta^2)^2}  \label{O4}
\end{eqnarray}
after a bit of algebra. We then recall that
\begin{equation}\label{O5}
    \int \frac{d^dq}{(2\pi)^d}\frac{1}{(q^2+\Delta)^n}=\frac{1}{(4\pi)^{d/2}}\frac{\Gamma(n-d/2)}{\Gamma(n)}(\Delta^2)^{d/2-n}\;,
\end{equation}
from which it follows that
\begin{equation}\label{O6}
    \int \frac{d^dq}{(2\pi)^d}\frac{q^2}{(q^2+\Delta)^n}=\frac{1}{(4\pi)^{d/2}}\frac{d}{2}\frac{\Gamma(n-d/2-1)}{\Gamma(n)}(\Delta^2)^{d/2-n+1}\;,
\end{equation}
and consequently also
\begin{equation}\label{O7}
    \int \frac{d^dq}{(2\pi)^d}\frac{q^4}{(q^2+\Delta)^n}=\frac{1}{(4\pi)^{d/2}}\frac{d(d+2)}{4}\frac{\Gamma(n-d/2-2)}{\Gamma(n)}(\Delta^2)^{d/2-n+2}\;.
\end{equation}
To obtain a finite result, we prefer to look at
\begin{eqnarray}\label{O8}
\frac{\p^2 F(k^2)}{(\p k^2)^2}&=& \frac{1}{4\pi}\int_0^1 dx \left[\left(12x^4-24x^3+14x^2-2x\right)\frac{1}{\Delta^2}+\left(8x^6-24x^5+25x^4-10x^3+x^2\right)\frac{k^2}{(\Delta^2)^2}+\right.\nonumber\\&&\left.+2x^4(1-x)^4\frac{k^4}{(\Delta^2)^3}\right]\;,
\end{eqnarray}
where we set $d=2$, as we are mainly interested in this case now.\\\\We consequently find
\begin{eqnarray}\label{O9}
\frac{\p^2 F(k^2)}{(\p k^2)^2}&=&\frac{1}{4\pi}\int_0^1 dx\left[\frac{-12x^2+12x-2}{k^2-is}+k^2\frac{8x^2-8x+1}{(k^2-is)^2}+2k^4\frac{x(1-x)}{(k^2-is)^3}\right]\;,
\end{eqnarray}
where we reintroduced $s=\frac{2x-1}{2x(1-x)}$, hence $x=\frac{-1+s+\sqrt{1+s^2}}{2s}$, which gives rise to
\begin{eqnarray}\label{O10}
\frac{\p^2 F(k^2)}{(\p k^2)^2}&=& \frac{1}{4\pi}\int_{-\infty}^{+\infty}\frac{ds}{2(1+s^2+\sqrt{1+s^2})}\left[-\frac{2}{s^2}(3+s^2-3\sqrt{1+s^2})\frac{1}{k^2-is}\right.\nonumber\\
&&\left.+\frac{k^2}{s^2}(4+s^2-4\sqrt{1+s^2})\frac{1}{(k^2-is)^2}+\frac{k^4}{1+\sqrt{1+s^2}}\frac{1}{(k^2-is)^3}\right]\;.
\end{eqnarray}
We first rewrite everything in terms of $k^2-is$ as follows
\begin{eqnarray}\label{O11}
\frac{\p^2 F(k^2)}{(\p k^2)^2}&=& \frac{1}{4\pi}\int_{-\infty}^{+\infty}\frac{ds}{2(1+s^2+\sqrt{1+s^2})}\left[-\frac{2}{s^2}(3+s^2-3\sqrt{1+s^2})\frac{1}{k^2-is}\right.\nonumber\\
&&\left.+\frac{1}{s^2}(4+s^2-4\sqrt{1+s^2})\frac{k^2-is+is}{(k^2-is)^2}+\frac{(k^2-is+is)^2}{1+\sqrt{1+s^2}}\frac{1}{(k^2-is)^3}\right]\nonumber\\
&=&\frac{1}{4\pi}\int_{-\infty}^{+\infty}\frac{ds}{2(1+s^2+\sqrt{1+s^2})}\left[-\frac{1}{s^2}\left(3+s^2-3\sqrt{1+s^2}\right)\frac{1}{k^2-is}\right.\nonumber\\
&&\left.+is\frac{-1+\sqrt{1+s^2}}{1+\sqrt{1+s^2}}\frac{1}{(k^2-is)^2}-\frac{s^2}{1+\sqrt{1+s^2}}\frac{1}{(k^2-is)^3}\right]\;.
\end{eqnarray}
Using two consecutive partial integrations, we can show that
\begin{eqnarray}\label{O12}
&&\int_{-\infty}^{+\infty}\frac{ds}{2(1+s^2+\sqrt{1+s^2})}\left[-\frac{1}{s^2}(3+s^2-3\sqrt{1+s^2})\frac{1}{k^2-is}\right]\nonumber\\
&=&-\int_{-\infty}^{+\infty}ds\left(\frac{-1+\sqrt{1+s^2}}{s^2}+\ln\left(1+\sqrt{1+s^2}\right)\right)\frac{1}{(k^2-is)^3}\;.
\end{eqnarray}
Similarly, partial integration  leads to
\begin{eqnarray}\label{O13}
&&\int_{-\infty}^{+\infty}\frac{ds}{2(1+s^2+\sqrt{1+s^2})}\left[is\frac{-1+\sqrt{1+s^2}}{1+\sqrt{1+s^2}}\frac{1}{(k^2-is)^2}\right]\nonumber\\
&=&\int_{-\infty}^{+\infty}ds \left[\frac{2}{1+\sqrt{1+s^2}}+\ln\left(1+\sqrt{1+s^2}\right)\right]\frac{1}{(k^2-is)^3}\;.
\end{eqnarray}
Hence, we can rewrite \eqref{O11} as
\begin{eqnarray}\label{O14}
\frac{\p^2 F(k^2)}{(\p k^2)^2}&=& \frac{1}{4\pi}\int_{-\infty}^{+\infty}ds\left[\frac{1}{2\sqrt{1+s^2}}\right]\frac{1}{(k^2-is)^3}
\end{eqnarray}
after simplification. We observe that there are no poles in the upper half $s$-plane for $k^2>0$, so we can fold our contour around the cut for $s\in[i\infty,i]$. With $s=i\tau$, we can write
\begin{eqnarray}\label{O15}
\frac{\p^2 F(k^2)}{(\p k^2)^2}&=& \frac{1}{4\pi}\int_{+\infty}^1id\tau\left[\frac{1}{-2i\sqrt{\tau^2-1}}\right]\frac{1}{(k^2+\tau)^3}+\frac{1}{4\pi}\int_1^{+\infty}id\tau\left[\frac{1}{2i\sqrt{\tau^2-1}}\right]\frac{1}{(k^2+\tau)^3}\nonumber\\
&=& \frac{1}{4\pi}\int_1^{+\infty}d\tau\frac{1}{\sqrt{\tau^2-1}}\frac{1}{(k^2+\tau)^3}\;.
\end{eqnarray}
We can now return to the original function $F(k^2)$. A first integration from 0 to $k^2$ gives
\begin{eqnarray}\label{O16}
\frac{\p F(k^2)}{\p k^2}-\left[\frac{\p F(k^2)}{\p k^2}\right]_{k^2=0}&=& \frac{1}{4\pi}\int_1^{+\infty}d\tau\frac{1}{-2\sqrt{\tau^2-1}}\frac{1}{(k^2+\tau)^2}-\frac{1}{4\pi}\int_1^{+\infty}d\tau\frac{1}{-2\sqrt{\tau^2-1}}\frac{1}{\tau^2}\;,
\end{eqnarray}
so that we get
\begin{eqnarray}\label{O17b}
&& F(k^2)-k^2\left[\frac{\p F(k^2)}{\p k^2}\right]_{k^2=0}-F(0)\nonumber\\&=& \frac{1}{4\pi}\int_1^{+\infty}d\tau\frac{1}{2\sqrt{\tau^2-1}}\frac{1}{k^2+\tau}+\frac{k^2}{4\pi}\int_1^{+\infty}d\tau\frac{1}{2\sqrt{\tau^2-1}}\frac{1}{\tau^2}- \frac{1}{4\pi}\int_1^{+\infty}d\tau\frac{1}{2\sqrt{\tau^2-1}}\frac{1}{\tau}\;.
\end{eqnarray}
Next, we shall also bring $G(k^2)$, give in eq.~\eqref{P2}, into a spectral form. Invoking the Feynman trick this time yields
\begin{eqnarray}
G(k^2)= \int_0^1 dx \int\frac{d^d q}{(2\pi)^d} \frac{M(k,q)}{(q^2+\Delta^2)^2}\;,  \label{P3}
\end{eqnarray}
with
\begin{eqnarray}
M(k,q) &=& k^2 q^2 \left( 1- \frac{1}{d}\right)\;.
\end{eqnarray}
As usual, we shall work in units $2\gam^2=1$. By using \eqref{O6}, we obtain the finite function
\begin{eqnarray}
\frac{\p^2 G(k^2)}{(\p k^2)^2}&=& \frac{1}{8\pi}\int_0^1 dx \left[\left(x^2 (1-x)^2\right)\frac{k^2}{(\Delta^2)^2} -\left(2 (1-x) x \right)\frac{1}{\Delta^2}\right]\;.
\end{eqnarray}
We substitute $x=\frac{-1+s+\sqrt{1+s^2}}{2s}$, so we obtain
\begin{eqnarray}
\frac{\p^2 G(k^2)}{(\p k^2)^2}&=& \frac{1}{8\pi}\int_{-\infty}^{+\infty}\frac{ds}{2(1+s^2+\sqrt{1+s^2})}\left[ \frac{k^2}{ (k^2 + i s)^2} -\frac{2}{ k^2 + i s} \right]\;.
\end{eqnarray}
Rewriting in terms of $k^2-is$ gives
\begin{eqnarray}\label{P4}
\frac{\p^2 G(k^2)}{(\p k^2)^2}&=& \frac{1}{8\pi}\int_{-\infty}^{+\infty}\frac{ds}{2(1+s^2+\sqrt{1+s^2})}\left[ \frac{-i s}{ (k^2 + i s)^2} -\frac{1}{ k^2 + i s} \right]\;.
\end{eqnarray}
Subsequently, using partial integration gives us the following identity
\begin{eqnarray}
\int_{-\infty}^{+\infty}\frac{ds}{2(1+s^2+\sqrt{1+s^2})}\left[ \frac{-i s}{ (k^2 + i s)^2}\right]&=&-\int_{-\infty}^{+\infty}ds\left[ \ln \left( \sqrt{s^2 +1 } + 1 \right) \right]\frac{1}{(k^2-is)^3}\;.
\end{eqnarray}
Analogically,
\begin{eqnarray}
\int_{-\infty}^{+\infty}\frac{ds}{2(1+s^2+\sqrt{1+s^2})}\left[\frac{-1}{(k^2-is)^2}\right]&=&\int_{-\infty}^{+\infty}ds \left[\sqrt{s^2 +1} -  \ln \left( \sqrt{s^2 +1 } + 1 \right)\right]\frac{1}{(k^2-is)^3}\;.
\end{eqnarray}
Therefore, eq.~\eqref{P4} becomes
\begin{eqnarray}
\frac{\p^2 G(k^2)}{(\p k^2)^2}&=& \frac{1}{8\pi}\int_{-\infty}^{+\infty}ds\frac{\sqrt{1+s^2}}{(k^2-is)^3}\,.
\end{eqnarray}
As before, we can fold our contour around the cut for $s\in[i\infty,i]$ and by setting $s=i\tau$, we find
\begin{eqnarray}
\frac{\p^2 G(k^2)}{(\p k^2)^2}&=& \frac{1}{4\pi}\int_{1}^{+\infty}d\tau \frac{\sqrt{\tau^2-1}}{(k^2+\tau)^3}\;,
\end{eqnarray}
and thus
\begin{eqnarray}
\frac{\p G(k^2)}{\p k^2}-\left[\frac{\p G(k^2)}{\p k^2}\right]_{k^2=0}&=& \frac{1}{4\pi}\int_1^{+\infty}d\tau\frac{\sqrt{\tau^2-1}}{-2}\frac{1}{(k^2+\tau)^2}-\frac{1}{4\pi}\int_1^{+\infty}d\tau\frac{\sqrt{\tau^2-1}}{-2}\frac{1}{\tau^2}\;.
\end{eqnarray}
Integrating a second time gives,
\begin{eqnarray}\label{laatst}
&& G(k^2)-G^2\left[\frac{\p G(k^2)}{\p k^2}\right]_{k^2=0}-G(0)\nonumber\\&=& \frac{1}{8\pi}\int_1^{+\infty}d\tau \sqrt{\tau^2-1}\frac{1}{k^2+\tau}+\frac{k^2}{8\pi}\int_1^{+\infty}d\tau\sqrt{\tau^2-1}\frac{1}{\tau^2}- \frac{1}{8\pi}\int_1^{+\infty}d\tau \sqrt{\tau^2-1} \frac{1}{\tau}\;.
\end{eqnarray}

\subsection{Using the Schwinger parametrization for $d=2$ and $d=4$}
As a check on our results \eqref{O19} and \eqref{P19}, we shall rederive these here using a different approach, and we shall simultaneously also treat the $d=4$ case. \\\\
To use the Schwinger parametrization,
\begin{align}\label{eq:Schwinger1}
 \frac{p^2}{p^4+\gamma^4}=\int_0^{\infty}d\al \cos(\gam^2 \al)e^{-\al p^2},\\
\label{eq:Schwinger2}
 \frac{\gam^2}{p^4+\gamma^4}=\int_0^{\infty}d\al \sin(\gam^2 \al)e^{-\al p^2},
\end{align}
we make the denominators real,
\begin{align}
 \langle& O^{(1)}_{\lambda\eta}(k) O^{(1)}_{\lambda\eta}(-k) \rangle =\mathcal{N}\int \frac{d^dp}{(2\pi)^d} 4\left(p^2(p-k)^2+(d-2)(p^2-p\,k)^2\right)\frac{p^2(p-k)^2+\theta^4}{(p^2-i\gam^2)((p-k)^4+i\gam^2)}  \nnnl
 &= 4\mathcal{N}  \int_0^\infty d\al\, d\be\, \cos\left(\gam^2(\al-\be)\right) \int\frac{d^dp}{(2\pi)^d} \left(p^2(p-k)^2+(d-2)(p^2-p\,k)^2\right) e^{-\al p^2 -\be (p-k)^2}\\
\langle & O^{(2)}_{\lambda\eta}(k) O^{(2)}_{\lambda\eta}(-k) \rangle  = \mathcal{N} \int \frac{d^dp}{(2\pi)^d} 32(k^2 p^2-(k\,p)^2)\frac{p^2(p-k)^2+\theta^4}{(p^2-i\gam^2)((p-k)^4+i\gam^2)} \;\nnnl
 &= 32\mathcal{N}   \int_0^\infty d\al\, d\be\, \cos\left(\gam^2(\al-\be)\right) \int\frac{d^dp}{(2\pi)^d}
(k^2 p^2-(k\,p)^2) e^{-\al p^2 -\be (p-k)^2}.
\end{align}
The Gaussian integrals can be evaluated by standard methods:
\begin{align}\label{eq:O1O1}
 \langle& O^{(1)}_{\lambda\eta}(k) O^{(1)}_{\lambda\eta}(-k) \rangle = 4\mathcal{N}  \int_0^\infty d\al\, d\be\, \cos\left(\gam^2(\al-\be)\right) (d-1)\nnnl
 &\times \left( \frac{d}{2}\left(\frac{d}{2}+1\right) (\al+\be)^{-d/2-2}+k^2 (\al^2+\be^2-d\,\al\,\be)(\al+\be)^{-d/2-3}+k^4\al^2\be^2(\al+\be)^{-d/2-4} \right),\\
\label{eq:O2O2}
\langle & O^{(2)}_{\lambda\eta}(k) O^{(2)}_{\lambda\eta}(-k) \rangle  = 32\mathcal{N} \int_0^\infty d\al\, d\be\, \cos \left(\gam^2(\al-\be)\right) (d-1) \frac{k^2}{2} (\al+\be)^{-d/2-1}.
\end{align}
For the remaining calculation it is useful to obtain a general expression for such integrals as appearing above. They can be derived along the lines of Section \ref{subsubsec:D4}, only that we consider now arbitrary powers of $\al$, $\be$ and $(\al+\be)$.\\\\
We define the expression $J$ as
\begin{align}
J(a,b,c,d)=(4\pi)^{-d/2}\int_0^\infty d\alpha \,d\beta \,\cos(\eta^2 (\al-\be)) \alpha^a \beta^b (\alpha+\beta)^c e^{-\frac{\alpha \beta}{\alpha+\beta}k^2}
\end{align}
and insert unity by means of \eref{swg-unity}. Then we rescale the variables $\alpha$ and $\beta$ by $\lambda$,
\begin{align}
J(a,b,c,d)=(4\pi)^{-d/2}\int_0^1 d\alpha \,d\beta \, \alpha^a \beta^b (\alpha+\beta)^c \int_0^\infty d\lambda\, \delta(\alpha+\beta-1) \lambda^{1+a+b+c} \,\cos(\lambda \,\eta^2 (\al-\be))  e^{-\frac{\alpha \beta}{\alpha+\beta}\lambda k^2},
\end{align}
evaluate the integrals over $\la$ and $\be$, and transform $\al$ to the variable $u=(2\al-1)/(2\al(1-\al))$:
\begin{align}
J(a,b,c,d)&=(4\pi)^{-d/2}\int_{-\infty}^\infty du \,\frac{1}{2}  \frac{1}{\sqrt{1+u^2}} \frac1{1+\sqrt{1+u^2}} \, \Gamma(2+a+b+c)\times \nnnl
 \times&\mathrm{Re} \Bigg\lbrace
 \left(\frac{-1+u+\sqrt{1+u^2}}{u}\right)^a \left(\frac{1+u-\sqrt{1+u^2}}{u}\right)^b 2^{2+c}  (1+\sqrt{1+u^2})^{2+a+b+c}
\left[ k^2-i\,2u\, \eta^2 \right]^{-2-a-b-c}
\Bigg \rbrace.
\end{align}
This integral is evaluated in the complex plane, where it possesses a branch cut from $u=i$ to $u=i \infty$. Setting $u=i\,y$ the integral $J$ is then given by
\begin{align}
J(a,b,c,d)&=(4\pi)^{-d/2}\Gamma(2+a+b+c)i\,2^{2+c}\int_1^\infty dy  \frac{1}{\sqrt{1-y^2}}  \times \nnnl
 \times&\mathrm{Re} \Bigg\lbrace
 (1+\sqrt{1-y^2})^{1+a+b+c}(-1+i\,y+\sqrt{1-y^2})^a (1+i\,y-\sqrt{1-y^2})^b  (i\,y)^{-a-b}
\left[ k^2+2y\, \gam^2 \right]^{-2-a-b-c}\Bigg \rbrace.
\end{align}
Now the physical cut from $k^2=-2\gam^2$ to infinity is manifest, but we still have to determine the spectral function. For this we rewrite the variable $y$ into an  angular variable $\chi$, $1/y=\cos \chi$, $0\leq\chi\leq\pi/2$,
\begin{align}\label{eq:J-def2}
J&(a,b,c,d)=(4\pi)^{-d/2}\,2^{2+c} \Gamma(2+a+b+c)i\int_0^{\pi/2} d\chi\, \frac{\sin \chi}{\cos^2\chi} \frac{1 }{i\sin\chi} \times \nnnl
 \times&\mathrm{Re} \Bigg\lbrace
 i^{-a-b}(\cos \chi+i\,\sin \chi)^{1+a+b+c} (-\cos \chi+i+i \sin \chi)^a (\cos \chi+i-i\sin \chi)^b (\cos \chi)^{-a-b-c}
\left[ k^2+2y\, \gam^2 \right]^{-2-a-b-c}\Bigg \rbrace=\nnnl
&=(4\pi)^{-d/2}\,2^{2+c} \Gamma(2+a+b+c) \times \nnnl
 \times&\int_0^{\pi/2} d\chi\,\mathrm{Re} \Bigg\lbrace
 i^{-a-b} e^{i\chi(1+a+b+c)} (-e^{-i\chi}+i)^a (e^{-i\chi}+i)^b (\cos \chi)^{-2-a-b-c}
\left[ k^2+2y\, \gam^2 \right]^{-2-a-b-c}\Bigg \rbrace,
\end{align}
and employ \eref{swg-ident} to obtain
\begin{align}\label{eq:J-def1}
J(a,b,c,d)&=(4\pi)^{-d/2}\,2^{2+c} \frac{1}{\Gamma(-1-a-b-c)} \int_{2\gam^2}^\infty d\tau \frac{1}{\tau+k^2}\times\nnnl
 \times&\int_0^{\varphi} d\chi\,\mathrm{Re} \Bigg\lbrace
 i^{-a-b} e^{i\chi(1+a+b+c)} (i-e^{-i\chi})^a (i+e^{-i\chi})^b (\cos \chi-\cos \varphi)^{-2-a-b-c} \tau^{-2-a-b-c}
\Bigg \rbrace,
\end{align}
where the order of integration was interchanged and $\varphi$ is defined by $\cos \varphi=2\theta^2/\tau$. Note that the external momentum $k^2$ in a coefficient of $J$ will turn into $-\tau$.\\\\
Eq. (\ref{eq:J-def1}) is valid for $-2-a-b-c\neq-1,-2,\ldots$. In our example there is indeed one case, where $-2-a-b-c=-1$ is realized. Here we use instead the intermediary expression \eref{eq:J-def2} of the derivation above. It leads directly to the spectral representation after replacing $\cos(\chi)$ by $2\gam^2/\tau$. Finally plugging eqs.~ (\ref{eq:J-def1}) and (\ref{eq:J-def2}) into eqs.~(\ref{eq:O1O1}) and (\ref{eq:O2O2}) yields
in two dimensions
\begin{align}
  \langle& O^{(1)}_{\lambda\eta}(k) O^{(1)}_{\lambda\eta}(-k) \rangle = 4\mathcal{N} \int_{2\gam^2}^\infty d\tau \frac{1}{\tau+k^2} \frac{\gam^4}{2\pi \sqrt{\tau^2-4 \gam^4}} \label{GZres1},\\
\label{GZres4}\langle & O^{(2)}_{\lambda\eta}(k) O^{(2)}_{\lambda\eta}(-k) \rangle  = 32\mathcal{N} \int_{2\gam^2}^\infty d\tau \frac{1}{\tau+k^2} \frac{\sqrt{\tau^2-4\gam^4}}{8\pi}.
\end{align}
These results are consistent with \eqref{O19} and \eqref{P19}. In four dimensions, we find
\begin{align}
\label{GZres2}  \langle& O^{(1)}_{\lambda\eta}(k) O^{(1)}_{\lambda\eta}(-k) \rangle = 12\mathcal{N} \int_{2\gam^2}^\infty d\tau \frac{1}{\tau+k^2} \frac{ \sqrt{\tau^2-4 \gam^4}(2\gam^4+\tau^2)}{32   \pi ^2 \tau} ,\\
\label{GZres3}\langle & O^{(2)}_{\lambda\eta}(k) O^{(2)}_{\lambda\eta}(-k) \rangle  = 96\mathcal{N} \int_{2\gam^2}^\infty d\tau \frac{1}{\tau+k^2} \frac{\left(\tau^2-4\gam^4\right)^{3/2}}{64\pi^2 \tau}\;.
\end{align}

\end{document}